\documentclass[aps,pra,twocolumn,nofootinbib,floatfix,longbibliography,superscriptaddress]{revtex4-2}

\usepackage{bm}

\usepackage[utf8]{inputenc}
\usepackage[english]{babel}
\babeladjust{autoload.bcp47 = on}
\usepackage[T1]{fontenc}
\usepackage{listings}
\usepackage{hyperref}
\usepackage{thm-restate}
\usepackage{thmtools}
\usepackage{comment}

\usepackage{amssymb}
\usepackage{amsthm}
\usepackage{bbm}
\usepackage{qcircuit}
\usepackage{adjustbox}
\usepackage{float}
\usepackage{caption}
\makeatletter \long\def\@makecaption#1#2{%
\par \begingroup \small \leftskip=0pt \rightskip=0pt \parfillskip=0pt plus 1fil \noindent \textbf{#1.} #2\par \endgroup } 
\makeatother

\expandafter\let\csname equation*\endcsname\relax

\expandafter\let\csname endequation*\endcsname\relax

\usepackage{natbib}
\usepackage{physics}
\usepackage{xcolor}
\usepackage{graphicx}
\usepackage{wrapfig}
\usepackage{qcircuit}
\usepackage{lipsum}

\usepackage{tikz}
\usepackage{tikz-cd}
\usetikzlibrary{positioning}
\tikzset{
  symbol/.style={
    draw=none,
    every to/.append style={
      edge node={node [sloped, allow upside down, auto=false]{$#1$}}}
  }
}

\usepackage{tcolorbox}
\tcbuselibrary{breakable}
\definecolor{exCol}{HTML}{fceeee}
\newtcolorbox[]{examplebox}[1][]{breakable,colback=exCol,colframe=white,#1}

\def\cP{{\mathcal P}}

\def\Tr{\text{Tr}}

\newcommand{\vleq}{\mathrel{\rotatebox{-90}{$\leq$\hphantom{tt}}}}

\theoremstyle{plain}

\begin{document}

\preprint{}

\title{Energetic Costs of Subspace Quantum Error Correction}

\author{Jakub Czartowski}
\affiliation{School of Physics, Trinity College Dublin, Dublin 2, Ireland}
\email{jakub.czartowski@tcd.ie}


\author{Felix C. Binder}
\affiliation{School of Physics, Trinity College Dublin, Dublin 2, Ireland}
\affiliation{Rinn Quantum, School of Physics, Trinity College Dublin}

	
\date{\today}

\begin{abstract}
    Quantum error correction acts as an entropy pump, transferring noise-induced uncertainty from a protected quantum system into syndrome information stored in an auxiliary memory. Repeated operation requires this memory to be cleared which unavoidably contributes to the energetic cost of error correction. Here, we characterise this contribution for subspace quantum error-correcting codes and identify how it depends on the joint structure of the code, the noise, and the representation of the retained syndrome information. Starting from the Knill--Laflamme conditions, we construct an effective syndrome state whose von Neumann entropy sets a lower bound on the ideal work required to maintain a reusable syndrome register. Projective syndrome readout generally generates additional entropy, and we quantify the resulting gap through measurement inefficiency. We then specialise to stabiliser codes under independent local Pauli noise and analyse two classical levels of syndrome representation. At the level of abstract error labels, degeneracies among single-qubit errors reduce the leading-order entropy of processed recovery labels. At the parity-check level, lower-weight checks reduce the marginal entropy generated by individual measurement outcomes in the low-noise regime. We identify the additional burden associated with retaining and separately erasing these outcomes as a bit-level inefficiency, and illustrate both costs for the five-qubit, Steane, generalised Shor, and rotated surface codes. Our results establish a hierarchy of syndrome-memory energetic costs and identify the code, noise, and measurement structures that control the ideal thermodynamic burden of subspace quantum error correction.
\end{abstract}

\maketitle

\section{Introduction}

\allowdisplaybreaks

Quantum error correction (QEC) is the central enabling ingredient for the future of scalable quantum information processing on noisy hardware. By distributing logical information across many imperfect physical degrees of freedom, QEC underpins the prospect of long-lived quantum memories, fault-tolerant quantum computation, and more broadly quantum technologies in which fragile quantum states must be protected against decoherence, control errors, and imperfect readout~\cite{Gottesman2009IntroQEC, LidarBrun2013QEC, spencer2026quantum, albert2026handbook}. Recent years have brought rapid progress in both logical-qubit experiments and fault-tolerant code design. Experimental demonstrations now include below-threshold surface-code memories with real-time decoding and improved logical error rates on trapped-ion processors~\cite{Google2024BelowThreshold, Paetznick2026, ransford2026helios0}. In parallel, high-threshold qLDPC codes and emerging planar constructions based on nearest-neighbour gates have strengthened the prospect of fault-tolerant quantum memories with substantially reduced qubit overhead~\cite{Bravyi2024HighThreshold, gu2026nearest0neighbour, nixon2026vine}. 
At the same time, QEC-based ideas are increasingly being exported beyond quantum computing, for example to quantum-enhanced imaging~\cite{Huang2022ImagingStars}, distributed entanglement-processing~\cite{Rengaswamy2024EntanglementPurification} or QEC-assisted axion detection~\cite{tan2026quantum}.

Because quantum information processing is realized in physical hardware, it is necessarily subject to thermodynamic constraints and costs. In particular, logically irreversible steps -- such as measurement recording, memory overwriting, and reset -- carry an entropy-related energetic cost that must ultimately be paid when the relevant information-bearing degrees of freedom are erased. First noted by Landauer~\cite{Landauer1961}, the unavoidable energetic cost of irreversible information processing is now well understood~\cite{Sagawa2009informationProcessing, Esposito2011, Reeb_2014, Faist2015, Timpanaro2020LandauerZero, 
Vu2022FiniteTimeLandauer, taranto2023LandauerNernst, Chattopadhyay2025LandauerReview}.

Quantum error correction is no exception. In a fully abstract description, encoding and recovery can be represented by unitary channels and so need not by themselves incur an energetic penalty stemming from irreversibility. Active QEC nevertheless requires logically irreversible ingredients: initialization, syndrome acquisition, classical recording of error information, decoder-side processing, and repeated reset of the corresponding memory. Earlier work has fruitfully related QEC to Maxwell's demon and Landauer erasure, in both classical and quantum settings~\cite{Vedral2000LandauerEC}. The complementary description of QEC as a thermal engine has been developed both for operator-based correction and for feedback-assisted, measurement-based protocols: environmental noise plays the role of a hot reservoir, while the auxiliary resources used for correction and reset act as a cold reservoir, and engine performance is linked to the recovery fidelity~\cite{Landi2020QECengines, Danageozian2022Triple}. What remains unresolved is a structural question: \emph{what is the minimal syndrome-related information that must be generated and eventually erased in realistic QEC, and how is the corresponding energetic cost constrained by the structure of the code and by the physical realization of syndrome extraction?}

In this work, we address this question in a layered way. We begin with the Knill--Laflamme (KL) conditions, which a subspace quantum error-correcting code must satisfy to correct a specified family of errors~\cite{KnillLaflamme2000}. We identify, at the most abstract level, the energetic cost stemming from syndrome acquisition needed for distinguishing correctable error sectors. We then specialize to stabiliser codes, where syndrome structure can be analysed systematically in terms of degeneracy classes. In combination with a noise model it yields an information-theoretic lower bound on the energetic cost associated with syndrome tracking, expressed in terms of the entropy of the corresponding syndrome representation. 

A central message of the paper is that this abstract syndrome-label entropy is generally only the beginning of the story. In a measurement-based stabiliser implementation, the physically primitive object generated by the hardware is a collection of ancilla readout bits produced by parity-check measurements. Any syndrome label is obtained only after classical postprocessing of that record. Consequently, the syndrome entropy relevant to thermodynamic accounting forms a hierarchy of non-decreasing costs: von Neumann entropy of effective pre-measurement syndrome state, the entropy of an abstract error label, the entropy of the raw but possibly correlated syndrome record, and the entropy of the uncompressed parity-bit stream need not and often do not coincide. In general, the amount of classical information that must be generated and temporarily stored can exceed what would be inferred from a naive evaluation of optimal KL error labels alone.

\vfill
\null

The paper is organized as follows. Section~\ref{sec:prelim} introduces QEC and thermodynamic preliminaries and identifies the fundamental energetic cost of irreversible information processing. Section~\ref{sec:cost_class} situates syndrome erasure within the broader energetic bookkeeping of QEC. Section~\ref{sec:label_track} develops an abstract description of syndrome tracking based directly on the Knill-Laflamme conditions. We show that energetic costs stemming from correctable noise arise from an effective syndrome state $\boldsymbol{\gamma}$ defined in terms of the interplay between noise characteristics and code subspace. Uncorrectable noise, on the other hand, corresponds to an input-dependent contribution to the effective syndrome state, which can be bounded in an input-independent way. Together, they impose fundamental lower bounds for energetic costs of QEC with syndrome measurements. Sec.~\ref{sec:stab_cost} specializes the results to stabiliser codes, where energetic costs decrease as code degeneracy increases, grouping different errors under the same correctable label, with leading-order advantage stemming from single-qubit errors. Section~\ref{sec:bit_level_cost} moves to the measurement-bit level, where erasure costs decrease together with check weights, pointing to a potential energetic advantage of qLDPC codes. Taken together, we find a hierarchy of energetic costs stemming from syndrome memory erasure, summarised schematically as:

\begin{widetext}
    \begin{equation}\label{eq:cost_hierarchy}
        \begin{array}{ccccccc}
        \text{Syndrome state }\boldsymbol{\gamma} &&
        \text{Error label }L && 
        \text{Measurement record }\vb{M} &&
        \text{Parity bits }\qty{M_j}_j
        \\\\
        S_{\rm vN}(\boldsymbol{\gamma})&
        \leq & H(L) &
        \leq & H(\vb{M}) &
        \leq & \sum_j H(M_j) \\
        \vleq &&\vleq &&\vleq &&\vleq\\
        W_{\boldsymbol{\gamma}} & \leq & W_{\mathrm{label}} & \leq & W_{\mathrm{record}} & \leq & W_{\mathrm{bit}}  \\
            | & & | & & & & | \\\hline
            \multicolumn{7}{|c|}{\text{Reduced by}} \\\hline
            \downarrow & & \downarrow &  & & & \downarrow \\
            \textbf{Code/Noise interplay} &&
            \textbf{Code degeneracy}
            &&&& \textbf{Low stabiliser weight} \\
            \text{Section \ref{sec:label_track}} &&
            \text{Section \ref{sec:stab_cost}} &&&& 
            \text{Section \ref{sec:bit_level_cost}}
        \end{array}
    \end{equation}
\end{widetext}
Sec.~\ref{sec:conclusions} concludes with a discussion of implications and limitations, and highlights directions for future work.

\section{Preliminaries}\label{sec:prelim}

In the following subsections we briefly recall notions that are necessary for a self-contained reading of this work, including subspace QECC, stabiliser formalism, and the minimal energy required for memory erasure. Additionally, we define a hierarchy of representations connected to error information, which will provide a recurring background to the subsequent considerations.

\subsection{Quantum error correction at the abstract level}

Let $\mathcal{C}$ be a code subspace
\footnote{Although our main discussion concerns subspace codes, many of the information-theoretic questions considered here admit analogues for subsystem codes, where only a protected subsystem must be recovered~\cite{Kribs06OQEC}. We do not pursue that generalization in the present work.} 
of a finite-dimensional Hilbert space $\mathcal{H}$, and let $P$ denote the projector onto~$\mathcal{C}$. A noise process $\mathcal{E}$ acting on the system with Kraus operators $\{E_i\}$ is correctable on $\mathcal{C}$ if there exists a recovery channel $\mathcal{R}$ such that
$
    \Tr_{\text{mem}}[\mathcal{R}\circ\mathcal{E}(\rho\otimes\op{0}_{\text{mem}})]=\rho
$
for all states $\rho$ supported on $\mathcal{C}$. The KL criterion states that this is possible if and only if
\begin{equation}
    \label{eq:KL_prelim}
    P E_i^\dagger E_j P = \alpha_{ij} P ,
\end{equation}
for some Hermitian matrix $\boldsymbol{\alpha}$~\cite{Nielsen_Chuang_2010,Gottesman2009IntroQEC}.

Operationally, recovery needs only to distinguish \emph{error sectors} whose action on the code is logically inequivalent, i.e., sitting in a distinct non-zero diagonal blocks of~$\boldsymbol{\alpha}$. Distinct Kraus operators need not correspond to distinct syndrome information: whenever different physical errors act identically on the code subspace up to a transformation that does not change the logical content, they belong to the same correctable class. This observation is the starting point for our later entropic analysis, where the minimal classical memory requirement is tied to the distinguishability of such error classes rather than to a naive count of Kraus labels.

For much of this paper we will exploit a basis adapted to the KL structure, in which the matrix $\boldsymbol{\alpha}$ is diagonal (or, more generally, block-diagonal in the presence of degeneracy). In that representation, correctable errors decompose into syndrome sectors that are orthogonal on the code space, making the relation between error distinguishability and classical syndrome information particularly
transparent. The intricacies of the more general non-diagonal case are addressed in Appendix~\ref{app:general}.

\subsection{Stabiliser codes and syndromes}

Stabiliser codes are an important, and indeed predominant subclass of QEC~\cite{spencer2026quantum, albert2026handbook}. We describe them here and will base most of our explicit analysis on them in the following.

For $n$ qubits, the Pauli group is
\begin{equation}
    \mathcal{P}_n = \{\pm 1,\pm i\}\cdot \{I,X,Y,Z\}^{\otimes n},
\end{equation}
where global phases will generally be ignored whenever they carry no operational significance. For a Pauli string $A=A_1\otimes\cdots\otimes A_n\in\mathcal{P}_n$, its weight $w(A)$ is the number of tensor factors $A_j$ that differ from the identity. 

A stabiliser code is specified by an Abelian subgroup $\mathcal{S}\subset\mathcal{P}_n$ that does not contain $-I$. The corresponding code space is the simultaneous $+1$ eigenspace
\begin{equation}
    \mathcal{C}(\mathcal{S})
    :=
    \qty{\ket{\psi}\in(\mathbb{C}^2)^{\otimes n}:
    S\ket{\psi}=\ket{\psi}\ \ \forall  S\in\mathcal{S}}.
\end{equation}
If $\mathcal{S}$ has $n-k$ independent generators, then $\mathcal{C}(\mathcal{S})$ has dimension $2^k$ and thus encodes $k$ logical qubits into $n$ physical qubits.

The normalizer of $\mathcal{S}$ in $\mathcal{P}_n$ is
\begin{equation}
    N(\mathcal{S})
    :=
    \{\,E\in\mathcal{P}_n : ESE^\dagger \in \mathcal{S}\ \ \forall S\in\mathcal{S}\,\}.
\end{equation}
Elements of $N(\mathcal{S})\setminus \mathcal{S}$ preserve the code space but act nontrivially on the encoded information, and therefore represent logical Pauli operators. By contrast, Pauli errors outside the normalizer move the state to a distinct syndrome sector and can in principle be detected by stabiliser measurements.

If $\mathcal{S}=\langle g_1,\dots,g_r\rangle$ is generated by $r=n-k$ independent commuting Pauli operators, then any Pauli error $E\in\mathcal{P}_n$ defines a binary syndrome vector
\begin{equation}
    \mathbf{s}(E)=(s_1(E),\dots,s_r(E))\in\{0,1\}^r,
\end{equation}
where
\begin{equation}
    s_j(E)=
    \begin{cases}
        0,& [E,g_j]=0,\\
        1,& \{E,g_j\}=0.
    \end{cases}
\end{equation}
Thus the syndrome records which generators commute or anticommute with the physical error. Errors differing by multiplication with a stabiliser element produce the same action on the code space and therefore belong to the same degenerate class. It is this merging of physically distinct errors into common syndrome classes that later gives rise to a reduction in syndrome entropy.

The distance of the stabiliser code is
\begin{equation}
    d
    :=
    \min\{\, w(A): A\in N(\mathcal{S})\setminus\mathcal{S}\,\}.
\end{equation}
Equivalently, $d$ is the minimum weight of a nontrivial logical Pauli operator. A code of distance $d=2t+1$ can correct all Pauli errors of weight at most $t$.

\subsection{Energetics of information processing}

Quantum error correction acquires a thermodynamic interpretation when a physical device records information about environmental noise and uses that information to protect an encoded state. The device may then be viewed as a version of Maxwell's demon~\cite{Maxwell1871,Vedral2000LandauerEC, Landi2020QECengines,Danageozian2022Triple}, with the syndrome memory playing the role of the demon's finite information register. During each correction cycle, this register stores syndrome information necessary to diagnose the realised error and determine the corresponding recovery operation. To sustain repeated operation, however, the register must eventually be restored to a standard clean state.

The work cost of restoring an information-bearing memory has been studied since Szilard's information-engine argument~\cite{Szilard1929} and Landauer's formulation of the erasure principle~\cite{Landauer1961}. For an otherwise energetically degenerate memory coupled to a thermal environment at temperature $T$, removal of an entropy $\Delta H_{\rm mem}\geq0$ requires work $\Delta W_{\rm er}$ satisfying~\cite{Parrondo2015}
\begin{equation}\label{eq:Landauer_prelim}
    \Delta W_{\rm er}\geq k_B T\Delta H_{\rm mem}.
\end{equation}
Here, $\Delta H_{\rm mem}$ denotes the entropy in nats, removed from the memory during the reset. In particular, if the register is restored to a pure standard state, $\Delta H_{\rm mem}$ coincides with the entropy of the syndrome record before erasure. The bound can be approached in the thermodynamically reversible limit.

Extensions of Landauer's Principle ask how closely the bound can be approached under physically realistic conditions by accounting for nonequilibrium memories, finite reservoirs, low temperatures, and finite-time operation~\cite{Sagawa2009informationProcessing,Esposito2011,Reeb_2014,Timpanaro2020LandauerZero,Vu2022FiniteTimeLandauer,taranto2023LandauerNernst,Chattopadhyay2025LandauerReview}. These refinements can modify the total physical work required for reset, but they do not alter the role of the record entropy in the original Landauer contribution which remains the fundamental limit. We therefore work in the ideal reversible-erasure setting and take the entropy of the retained error-correction record as the information-theoretic measure of its minimum energy cost for reset, saturating the bound in \eqref{eq:Landauer_prelim}.

\subsection{Syndrome representations used in this work}

A recurring theme in this work is that there are several inequivalent ways of representing syndrome information, each associated with a different entropic cost.

\begin{figure*}
    \centering
    \includegraphics[width=\linewidth]{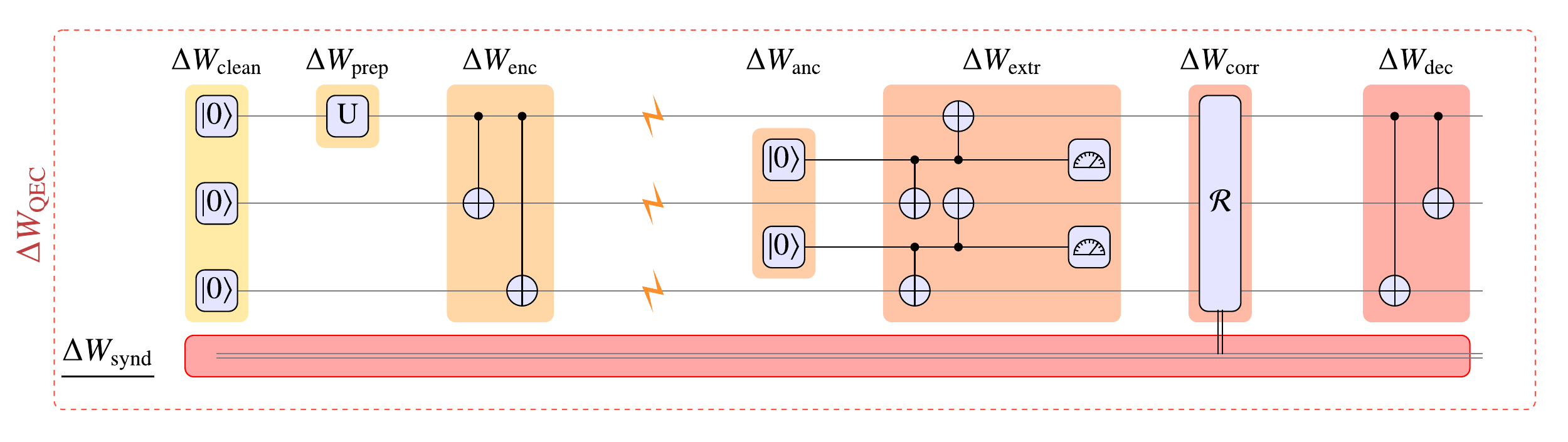}
        \caption{\textbf{Different costs of QEC:} A single round of repetition code without computation, illustrating different potential sources of energetic costs in quantum error correction, with single wires representing quantum systems and double wires classical data. In this work we focus on the energetic cost $\Delta W_\text{synd}$ of syndrome data measurement and erasure, which can be seen as demonstrably irreversible.}
    \label{fig:cost_pictorial}
\end{figure*}

\begin{enumerate}
    \item \textbf{Coherent syndrome information.} Prior to measurement, information about the realised noise is encoded in auxiliary quantum degrees of freedom, described by an effective syndrome state $\boldsymbol{\gamma}$, arising from interplay between code and noise. $\boldsymbol{\gamma}$~may contain coherences inaccessible to the subsequent syndrome measurement.

    \item \textbf{Processed syndrome labels.}
    Following measurement and classical postprocessing, the syndrome information may be reduced to an error class identifier $L$, or equivalently a decoder label, sufficient to select the corresponding recovery action.

    \item \textbf{Raw one-round syndrome records.}
    In a measurement-based implementation, syndrome measurement first generates a binary vector of parity-check outcomes,
    \begin{equation*}
        \mathbf{M}=(M_1,\dots,M_r)\in\{0,1\}^r,
    \end{equation*}
    typically by measuring ancilla qubits coupled to the stabiliser generators. The processed label $L$ is then obtained from $\mathbf{M}$ by classical postprocessing.

\end{enumerate}

These representations are related but not interchangeable. 
One of the central aims of this work is to clarify how these different levels of representation translate into different lower bounds on the thermodynamic cost of syndrome tracking and reset. Sections~\ref{sec:label_track} and~\ref{sec:stab_cost} focus primarily on processed syndrome labels $L$, Sec.~\ref{sec:bit_level_cost} on raw parity-measurement records $\mathbf{M}$.

\section{Full Cycle Cost of Quantum Error Correction}
\label{sec:cost_class}

Before turning to the syndrome-related contribution that forms the main scope of this work, it is useful to place it within the broader energetic tally of an active quantum error-correction protocol.
Even if ideal encoding, decoding, and recovery are represented by reversible unitary maps, a realistic QEC cycle involves several additional steps.
Some of these contributions depend strongly on the implementation, whereas others admit a more universal information-theoretic interpretation. The purpose of the present section is to separate these contributions at a high level and identify the syndrome-related part that will be studied in what follows.

\medskip
\noindent
\textbf{\textit{One-off initial and final costs}}
--
At the beginning and the end of the protocol one typically incurs several one-off energy costs:
\begin{enumerate}
    \item $\Delta W_{\mathrm{clean}}$: cost of preparing the physical registers
    in a clean initial state. If no knowledge of the prior state is assumed, this step itself involves energetic cost.
    
    \item $\Delta W_{\mathrm{prep}}$: cost of preparing the input state \mbox{$\ket{\psi} = U\ket{0}$} to be protected. In practice this includes implementation-dependent gate costs and preparation noise.
    
    \item $\Delta W_{\mathrm{enc}} + \Delta W_{\mathrm{dec}}$: cost of encoding and decoding which, as in the previous step, may incur implementation-dependent gate costs.
\end{enumerate}

\medskip
\noindent
\textbf{\textit{Repeated recovery}}
--
The main recurring contribution comes from repeated application of the recovery cycle. At a broad level, this cost may be decomposed into several parts:
\begin{enumerate}
    \setcounter{enumi}{3}
    \item $\Delta W_{\mathrm{anc}}$: cost of ancilla preparation or reset which involves returning syndrome qubits to the clean state required for the next round.
    
    \item $\Delta W_{\mathrm{extr}}$: cost of syndrome extraction, including gate operations, measurement, and readout.  In an ideal reversible model, acquisition need not itself incur energetic cost if a clean record register is available.
    
    \item $\Delta W_{\mathrm{synd}}$: cost of storing and erasing syndrome information. Here the relevant record may be represented as raw measurement bits, as a processed syndrome label, or as some decoder state retained between extraction and correction.
    
    \item $\Delta W_{\mathrm{corr}}$: cost of applying the corresponding correction or, equivalently, of updating the relevant Pauli frame. In the idealized description this need not itself contribute a fundamental irreversible cost, but its realization may still be dissipative.
\end{enumerate}

It is worth noting that the unitary gates employed in QEC are reversible and therefore entail no fundamental requirement for energy dissipation. Their physical implementation may nevertheless require energetic resources for control and may be subject to energy–precision trade-offs~\cite{Stevens2025}, which lie beyond the scope of this work.

Taken together, the total cost of operating the code is therefore
\begin{equation}
\begin{aligned}
    \Delta W_{\mathrm{QEC}} =\;&
    \Delta W_{\mathrm{clean}}
    + \Delta W_{\mathrm{prep}} \\
    &+ \Delta W_{\mathrm{enc}}
    + \Delta W_{\mathrm{dec}}
    + \Delta W_{\mathrm{rec}},
\end{aligned}
\end{equation}
with
\begin{equation}
    \Delta W_{\mathrm{rec}} =
    \Delta W_{\mathrm{anc}}
    + \Delta W_{\mathrm{extr}}
    + \Delta W_{\mathrm{synd}}
    + \Delta W_{\mathrm{corr}}.
\end{equation}

In this general tally (see Fig.~\ref{fig:cost_pictorial}), several terms depend strongly on the specific hardware and cannot be bounded sharply without an explicit implementation model. The main exception is the syndrome-related classical memory: once syndrome information has been generated and stored, making that memory available again requires a logically irreversible reset, and it is this contribution that incurs a fundamental energetic cost. For that reason, in what follows we focus primarily on $\Delta W_{\mathrm{synd}}$ and use entropy production in rapid-access syndrome memory (RASM) as the main proxy for the fundamental thermodynamic burden of repeated error correction.

A final remark concerns classical postprocessing. In a realistic fault-tolerant architecture, decoding and processing syndromes requires nontrivial classical resources~\cite{Roffe2020, Ueno2021QECOOL,Barber2025, Caune2026RealTime}, which necessarily consume energy. In the present work we do not attempt to include the full cost of external classical hardware, and instead restrict attention to RASM, presumably positioned close to the error-corrected quantum memory, whose reset is directly tied to the operation of the error-corrected quantum device itself.

\section{Cost of Syndrome Tracking}
\label{sec:label_track}

In what follows we consider energetic costs of subspace quantum error correction under the standard assumption of orthogonal error subspaces, with detailed derivations deferred to Appendix~\ref{app:effective_syndrome_derivation}. We begin with the diagonalised nondegenerate case of the KL condition~\eqref{eq:KL_prelim}. We define a set of operators $\qty{F_i}_{i=1}^r$ which provide the basis for syndrome detection, satisfying
\begin{equation} \label{eq:diag_KL}
    P F_i^\dagger F_j P = \delta_{ij} P.
\end{equation}
We work under the assumption that detection of syndromes is done by performing a measurement corresponding to projections $P_i = U_i P U_i^\dagger$ onto the rotated copies of the code, with relation $F_i P = U_i P$ proceeding from polar decomposition. Thus, the recovery map $\mathcal{R}$ is described by Kraus operators $R_i = U_i^\dagger P_i \otimes \op{i}{0}$.

Since the KL condition effectively associates the code $\mathcal{C}$ with a correctable vector subspace $\mathcal{F} = \operatorname{span}(F_i)$, we may work in any operator basis, in particular $\widetilde{F}_j = \sum_{i=1}^r V_j^i F_i$ with $V$ unitary, thus preserving $P\widetilde{F}_i^\dagger\widetilde{F}_jP = \delta_{ij}$. 
\smallskip

We first focus on the correctable noise \mbox{$\mathcal{E}_{\text{corr}}$} with  restricted Kraus operators of the form \mbox{$K_a = \sum_{i=1}^r c_a^i F_i$}. By identifying $\boldsymbol{\gamma}_{i'i} = \sum_a {c^*}_a^{i'}c_a^i$ as the effective density matrix of the syndromes we find that the probability distribution of syndromes is given by
\begin{align}\begin{split}
    \Tr_{\text{sys}}\left[ \mathcal{R}\circ\mathcal{E}(\rho\otimes\op{0})\right] &= {} \\\qquad \sum_{i,i',j}{V^T}_{i'}^{j} \boldsymbol{\gamma}_{i'i} {V^\dagger}_i^j \op{j} &= {}  \\
    \qquad \ev{(V\boldsymbol{\gamma} V^\dagger)}{j} \op{j} &=: \mathcal{D}_V(\boldsymbol{\gamma}) \label{eq:eff_dens_mat}
\end{split}\end{align}
as projection of the rotated state $V \boldsymbol{\gamma} V^\dagger$ onto the computational basis. Thus, work $\Delta W_{\mathrm{synd}}$ needed to erase syndrome entropy is lower-bounded by optimal measurement saturating the von Neumann entropy of the syndrome density matrix,
\begin{equation}
    S_{\text{vN}}(\boldsymbol{\gamma}) \leq H[\mathcal{D}_V(\boldsymbol{\gamma})] \equiv \kappa_{\mathrm{meas}} S_{\text{vN}}(\boldsymbol{\gamma}) \leq \frac{\Delta W_{\rm synd}}{k_B T} 
\end{equation}
where $H[\mathcal{D}_V(\boldsymbol{\gamma})]$ is the standard Shannon entropy and $S_{\text{vN}}(\boldsymbol{\gamma})$ is the von Neumann entropy of the effective syndrome state. Additionally, we identify $\kappa_{\mathrm{meas}} = \frac{H[\mathcal{D}_V(\boldsymbol{\gamma})]}{S_{\text{vN}}(\boldsymbol{\gamma})}$ as the inefficiency arising from irreversibility of measurement process.
Saturation can also be achieved by reversible  unitary implementation of the recovery map $\mathcal{R}$ without intermediate measurement (see Appendix \ref{app:effective_syndrome_derivation}).

This highlights that the entropy production of quantum error correction is connected directly to the interplay between the selected syndromes and noise decomposition, with the entropy-optimal syndromes selected in such a way as to remove any coherences from the effective state of the syndrome qubits. 

In Appendix~\ref{app:general} we extend the discussion to a more general case of non-diagonal $\boldsymbol{\alpha}$ and discuss trade-offs stemming from non-orthogonal projectors, which necessitate the introduction of a failure probability.

\medskip
Let us now proceed to consider the effect of uncorrectable noise on the syndrome entropy. For this purpose, we will consider the extension of the error basis to $\qty{F_i}$ such that $i\leq r$ label errors that satisfy the KL conditions of the form~\eqref{eq:diag_KL}, and additionally
\begin{equation}
        P F_i^\dagger F_j P = \Delta_{ij}
\end{equation}
for $i\leq r < j$, where $\Delta_{ij}\neq P$ are the logical errors introduced by uncorrectable noise, and transform under change of syndrome basis as $\sum_{k} {V^\dagger}^k_i\Delta_{kj} := \widetilde{\Delta}_{ij}$. 
With this, we shift focus to the uncorrectable part of the noise $\mathcal{E}_{\mathrm{unc}}$ described by Kraus operators decomposable into purely uncorrectable errors\footnote{More generally, an individual Kraus operator may have support in both the correctable and uncorrectable error sectors, $K=K_{\mathrm{cor}}+K_{\mathrm{unc}}$. Its action then generates the additional cross terms $K_{\mathrm{cor}}\rho K_{\mathrm{unc}}^\dagger$ and $K_{\mathrm{unc}}\rho K_{\mathrm{cor}}^\dagger$, which we omit to keep the presentation concise. The analysis can be extended to include
these contributions.}, $K'_a = \sum_{j>r} {c'}^j_a F_j$. 
With this, contribution to the probability of syndromes arising from purely uncorrectable noise is given by
    \begin{align}\begin{split}
            \Tr_{\text{sys}}\qty[ \mathcal{R}\circ\mathcal{E}_{\text{unc}}(\rho\otimes\op{0})] = {} \\
           \qquad \sum_{a, j} \Tr\qty[A_{aj}^{(V)} \rho A_{aj}^{(V)\dagger}]  \op{j}
         \label{eq:uncorr_prob}
    \end{split}\end{align}
where $A_{aj}^{(V)} = \sum_{k>r}\sum_{\ell =1}^r {c'}_a^k {V^\dagger}_j^\ell  \Delta_{\ell k}$.
Thus, the contribution to syndrome label $j$ is bounded as

\begin{equation}\label{eq:uncorr_prob_bound}
    {\tiny p_{\text{unc}}(j) = \sum_a \norm{A_{aj}^{(V)} \sqrt{\rho}}_2^2 \leq \sum_a \norm{A_{aj}^{(V)}}_\infty^2}
\end{equation}
where we used $\norm{\sqrt{\rho}}_2^2 = \Tr(\rho) = 1$. One may further obtain bounds in terms of logical errors $\Delta_{ij}$ and noise coefficients ${c'}^{i}_a$ by application of properties of $\infty$-norm and Cauchy-Schwarz inequality, which yield

\break
    \begin{subequations}
        \label{eq:uncorr_prob_bound_hier}
        \begin{align}
                 p_{\text{unc}}(j) &\leq \sum_a \sum_{\ell=1}^r  \qty(\sum_{i>r}\abs{{c'}_a^i}\norm{\Delta_{\ell i}}_\infty)^2\\
                 &\leq \sum_a \sum_{i>r} \abs{{c'}_a^i}^2 \sum_{i>r}\sum_{\ell=1}^r \norm{\Delta_{\ell i}}_\infty^2.
        \end{align}
    \end{subequations}
Each of these provides a basis- and index-independent upper bound of the uncorrectable noise contribution to the syndrome probability distribution. 

An alternative interpretation follows from viewing the uncorrectable errors as resulting in a state-conditioned contribution to the syndrome matrix.  To see this, we can re-express Eq.~\eqref{eq:uncorr_prob} as
\begin{equation} \label{eq:eff_uncorr_dens_mat}
p_{\text{unc}}(j)=\sum_{\ell ,\ell' =1}^r V^j_\ell \Gamma_{\ell \ell'}(\rho)(V^*)_{\ell'}^j=\ev{V \Gamma(\rho)V^\dagger}{j},
\end{equation}
where
\begin{equation}
\Gamma_{\ell \ell'}(\rho):=\sum_{i,i'>r}\boldsymbol{\gamma}_{ii'}'\Tr[\Delta_{\ell i}\rho \Delta_{\ell'i'}^\dagger],
\end{equation}
and $\boldsymbol{\gamma}'_{i'i}:=\sum_a  {c'^*}_a^{i'}{c'}_a^i$ as in the correctable case.
The matrix $\Gamma(\rho)$ therefore plays the role of an effective syndrome matrix contribution from the uncorrectable sector.  However, unlike in the correctable KL sector, $\Gamma(\rho)$ is not determined solely by properties of the noise encoded in $\boldsymbol{\gamma}'$. The operators $\Delta_{lk}$ act non-trivially on the code space and hence make the syndrome statistics explicitly dependent on the logical input state $\rho$.  Thus $V^\dagger \Gamma(\rho)V$ may still be interpreted as a contribution to the effective syndrome density matrix, but only in a state-conditioned sense: it is obtained by evaluating the uncorrectable operator structure on the code state, rather than representing an autonomous syndrome state independent of $\rho$. As a result, there is no state-independent error basis allowing for minimising the energetic costs of error correction once uncorrectable errors are taken into account.

\medskip

Note that the analysis presented in this section parallels the general arguments made in~\cite{Vedral2000LandauerEC}, where entropic consequences of QEC have been established without demonstrating relation between selected QEC scheme and its energetic costs. In contrast, we develop specific relations between the subspace defining the code, the error subspace defining the correctable errors satisfying KL conditions, and the structure of the noise itself, together giving rise to the effective syndrome state~$\boldsymbol{\gamma}$, which provides explicit lower bounds on the energetic costs of measured syndrome memory erasure.

\section{Costs of stabiliser codes}
\label{sec:stab_cost}

In this section we turn the attention from abstract QEC to qubits and stabiliser codes.
We use a standard model for Markovian qubit noise given by the generic noise channel
\begin{equation} \label{eq:1_qub_noise_uniform}
    \mathcal{E}(\rho) = \qty(1-p_x-p_y-p_z)\rho + \sum_{i\in\qty{x,y,z}}p_i \sigma_i\rho\sigma_i.
\end{equation}
Note that, in this case, the stabiliser measurements and the noise are expressed in the same Pauli basis. Consequently, the effective syndrome state $\boldsymbol{\gamma}$ is diagonal in the measurement basis, and the measurement inefficiency is $\kappa_{\mathrm{meas}}=1$. 

More generally, one may consider a mismatch between the syndrome and noise axes, with the latter defined by the rotated Pauli operators $\sigma'_i = \sum_j R_{ij}\sigma_j$ where $R$ is an orthogonal matrix. The corresponding noise channel is
\begin{equation}
    \label{eq:1_qub_noise_uniform_rotated}
    \begin{aligned}
        \mathcal{E}(\rho)
        ={}& \qty(1-p_x-p_y-p_z)\rho \\
        &+ \sum_{i\in\qty{x,y,z}} p_i
        \sum_{j,k} R_{ij}R_{ik}\sigma_j\rho\sigma_k .
    \end{aligned}
\end{equation}
Measurement in the original syndrome basis then yields the probabilities $p'_j = \sum_i R_{ij}^{\,2}p_i$. Since the matrix with entries $R_{ij}^{\,2}$ is doubly stochastic, the resulting probability vector is more mixed, and hence
\begin{equation}
    H\qty(\mathcal{D}_V(\boldsymbol{\gamma}))
    = H\qty(\{p'_j\}_j)
    \geq H\qty(\{p_i\}_i)
    = S(\boldsymbol{\gamma}) .
\end{equation}
The inequality is strict whenever the rotation produces a nontrivial mixing of unequal noise probabilities, in which case $\kappa_{\mathrm{meas}}>1$. At the same time, the terms with $j\neq k$ generate off-diagonal contributions to $\boldsymbol{\gamma}$ and therefore constitute its coherent sector.

For simplicity we focus on the symmetric unbiased case $p_x=p_y=p_z=p$, which corresponds to the usual depolarising channel\footnote{Shift to biased noise is straightforward and does not alter the general scaling conclusions of this section.}. This translates to multiqubit noise given by
\begin{equation} \label{eq:n_qub_noise_uniform}
    \mathcal{E}^{\otimes n}(\rho) = \sum_{A \in\cP_n/\qty{\pm1,\pm i}} (1-3p)^{n-w(A)}p^{w(A)} A\rho A^\dagger,
\end{equation}
with global phase factored out to avoid overcounting. 

Let us now consider a stabiliser code with parameters $[[n,k,2t+1]]$ and define the set $\cP_{n,m} := \qty{A: w(A) = m} \subset \cP_n$ as the set of Pauli strings of weight $m$. For the code with the distance $2t+1$ a guaranteed set of correctable errors $\cP_{n,i}$ is given by all Pauli strings of weight at most $t$ composed with arbitrary element of the underlying stabiliser group, $\bigcup_{i=0}^{t} \cP_{n,i} \subset \qty{SA: S\in\mathcal{S}, w(A)\leq t}$. 

For sufficiently small $p$ satisfying
\begin{equation}\label{eq:cutoff_condition}
     \begin{aligned}
     (1-3p)^{n-1} 3p \binom{n}{1} & \gg  (1-3p)^{n-2} (3p)^{2}\binom{n}{2} \\
     \Rightarrow 3p&\ll \frac{2}{n-1}.
     \end{aligned}
\end{equation}
two-qubit errors are much less likely than single qubit errors, $E_i\in\cP_{n,1}$. In this limit we identify simple relation between single-qubit errors: $E_i, E_j\in\cP_{n,1}$ acting on two different qubits are equivalent in terms of identical syndrome, $E_i\sim E_j$, if and only if $E_i^\dagger E_j\in\mathcal{N}(\mathcal{S})$. We denote the equivalence classes as~$[E_i]$.  All such degeneracies are therefore encoded by the weight-two elements of the normaliser, $\cP_{n,2}^{\mathcal{N}} := \mathcal{N}(\mathcal{S})\cap\cP_{n,2}$, each of which connects two single-qubit errors within the same equivalence class.
Thus, in the regime where single qubit errors are much more likely than other errors -- i.e., $p \ll \frac{2}{3(n-1)}$ (Eq.~\eqref{eq:cutoff_condition}) -- the entropy of the resulting syndrome probability $\vb{p}_{\text{synd}}$ is
    \begin{equation} \label{eq:entropy_stab}
    \begin{aligned}
        H(\vb{p}_{\mathrm{synd}})
        = &
        -p_0 \log p_0
        - (3n)p \log p \\
        & - p \sum_{[E_i]} \abs{[E_i]} \log \abs{[E_i]}
        + \mathcal O(p^2),
    \end{aligned}
    \end{equation}
with $p_0 = (1 - 3p)^n$ the probability of no physical error. As a consequence, leading contribution to energetic costs at the level of syndrome labels benefits from single-qubit error degeneracy and can be read off immediately from weight-2 members of the stabiliser group.

While similar arguments could be formulated for errors of higher weights, $w(E)>1$, in that case, the syndrome degeneracy does not reduce to a straightforward structure as described above. For this reason, we defer detailed considerations of higher-weight sectors to Appendix~\ref{app:higher-weight}.

We illustrate the energetic cost of syndrome erasure at the label level with several example applications of the entropy formula~\eqref{eq:entropy_stab}.

\begin{examplebox}[boxsep=0pt,left=1em,right=1em]     
    \phantomsection \label{ex1}   
    \begin{center}
        {\fontsize{10pt}{12pt}\selectfont {\bf Example 1: Five-qubit perfect code and Steane code}}
    \end{center}

        The five-qubit code $[[5,1,3]]$~\cite{Laflamme1996} and Steane code $[[7,1,3]]$~\cite{CSSSteane} are examples of non-scalable quantum error correction codes. While the Steane code is based on a Hamming code $[7,4,3]$ used in a CSS construction, the five-qubit code is a minimal code saturating the quantum Hamming bound. Since neither code is degenerate on a single-qubit error level, the expressions for syndrome entropy under symmetric depolarising noise reduce to
        \begin{subequations} \label{eq:ent_perf_stean_lab}
            \begin{align}
                \begin{aligned}
                    H_{\mathrm{label},\text{5-q}}(p) = & -(1-15p)\log(1-15p) \\
                    & - 15 p\log p  + \mathcal{O}(p^2\log p) \\
                    \approx & 15p(1-\log p)
                \end{aligned} \\
                \begin{aligned}
                    H_{\mathrm{label},\text{Steane}}(p) = & -(1-21p)\log(1-21p) \\
                    & - 21 p\log p  + \mathcal{O}(p^2\log p) \\
                    \approx & 21p(1-\log p)
                \end{aligned}        
            \end{align}
        \end{subequations}
        with $n=5, 7$ respectively.
\end{examplebox}

\begin{examplebox}[boxsep=0pt,left=1em,right=1em]
    \phantomsection \label{ex2}
    \begin{center}
        {\fontsize{10pt}{12pt}\selectfont {\bf Example 2: $n^2$-qubit Shor codes}}
    \end{center}
        Originally introduced by Peter Shor in~\cite{Shor1995ShorCode}, the $[[9,1,3]]$ Shor code can be seen as a concatenation of a $Z$-repetition code on top of an $X$-repetition, which provides a realization of a wide family of CSS codes~\cite{CSSCalderbankShor, CSSSteane}. 

        Considering a double index for qubits, which are understood to be arranged in a square grid, the set of stabilisers for this code is given by pairwise stabilisers $\qty{Z_{(i,j)}Z_{(i,j+1)}}_{i=1, j=1}^{n,n-1}$ second-layer stabilisers $\qty{\prod_{j=1}^n X_{(i,j)}X_{(i+1,j)}}_{i=1}^{n-1}$. Restricting ourselves to single-qubit errors as before, all $X$ and $Y$ errors generate distinct syndromes, while $Z$ errors exhibit $n$-fold degeneracy. This results in reduction of entropy under symmetric depolarising noise,
        \begin{equation}\label{eq:ent_shor_lab}
            {\small\begin{aligned}
                H_{\mathrm{label},\text{Shor}}(p) = & -(1-3n^2p)\log(1-3n^2p) \\ & - 3n^2 p\log p - p n^2\log n + \mathcal{O}(p^2\log p) \\
                \approx & n^2p\qty(3- 3\log p - \log n).
            \end{aligned}}
        \end{equation}
       Here, the advantage due to degeneracy -- $pn^2\log n$ -- scales faster than the number of physical qubits.
\end{examplebox}
\phantomsection \label{ex3}
\begin{examplebox}[boxsep=0pt,left=1em,right=1em]     
    \begin{center}
        {\fontsize{10pt}{12pt}\selectfont {\bf Example 3: Rotated surface code}}
        \includegraphics[width=0.45\linewidth]{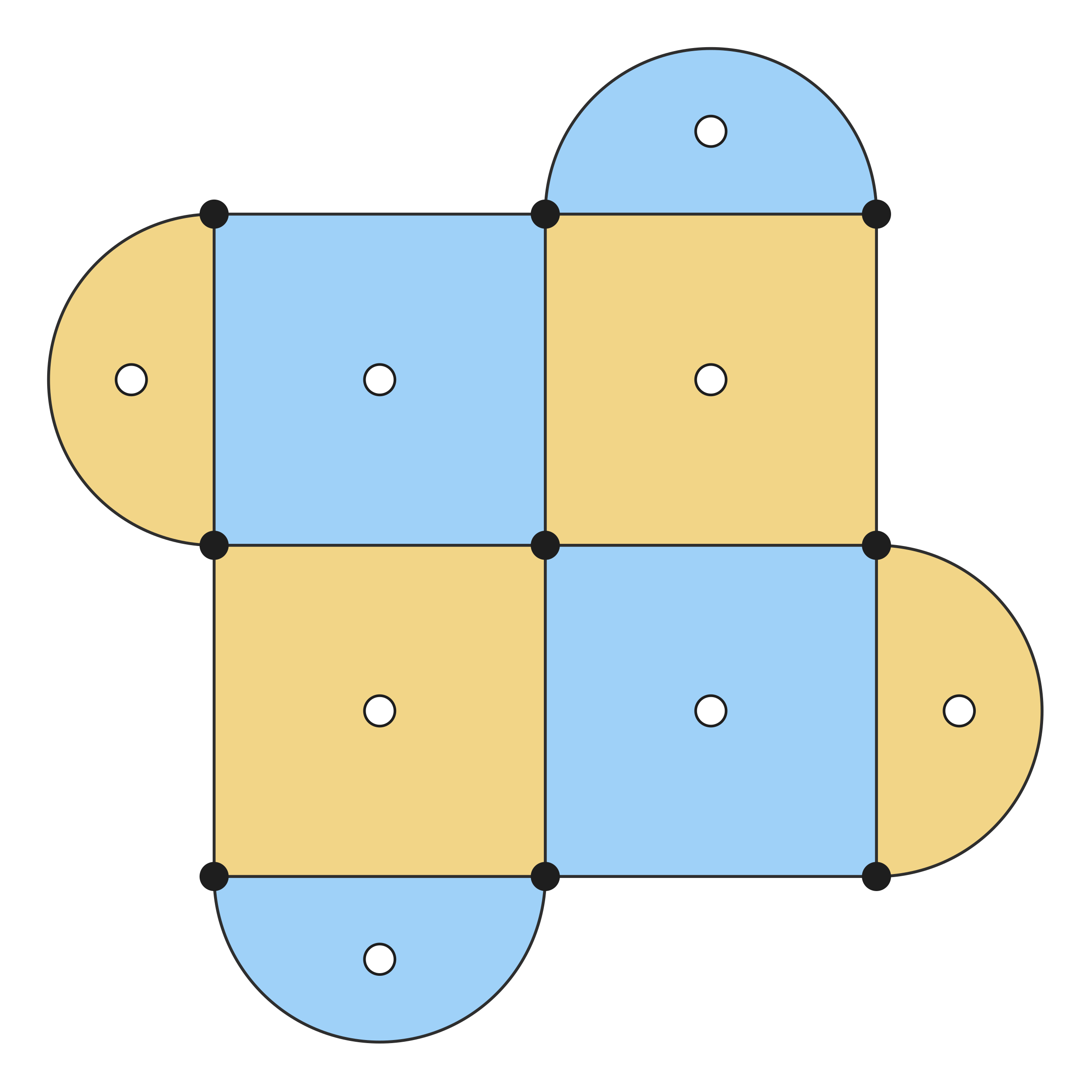}
    \captionof{figure}{$3\times 3$ patch of the rotated surface code $[[9,1,3]]$. Black (white) circles denote data (syndrome) qubits. Orange and blue plaquettes indicate the supports of the $X$- and $Z$-type stabilisers, respectively. Each stabiliser acts on the data qubits at the corners of its plaquette and is measured via the syndrome qubit at its centre.}\label{fig:patch_33}
    \end{center}

        The rotated surface code provides an example of a topological code, which uses a square lattice of side $n$ to encode a single logical qubit in a stabilizer code $[[n^2,1,n]]$~\cite{bravyi1998quantum, Bombin2007optimal2dcodes, Horsman2012}. The $n=3$ case is illustrated in Fig.~\ref{fig:patch_33}. The stabilisers for the code are represented by plaquettes with orange plaquettes corresponding to an $X$-type stabiliser measured jointly on all the corners and similarly with $Z$-type for blue plaquettes. They can be divided into $(n-1)^2$ bulk-type plaquettes of Pauli weight 4, and the remaining $2(n-1)$ being of edge-type of weight 2. Each weight-two boundary stabiliser pairs two same-type single-qubit errors with an identical syndrome. Altogether, the boundary checks produce $2(n-1)$ two-element degeneracy classes. This results in syndrome entropy
        \begin{equation}\label{eq:ent_surf_lab}
        {\small
            \begin{aligned}
                H_{\mathrm{label},\text{surf}}(p) = & -(1-3n^2p)\log(1-3n^2p) \\ & - 3n^2 p\log p - 4p(n-1)\log 2 \\
                & + \mathcal{O}(p^2\log p) \\
                \approx & 3n^2p\qty(1-\log p) - 4p(n-1)\log 2.
            \end{aligned}}
        \end{equation}
        where the degeneracy advantage scales only with the linear size of the lattice.
\end{examplebox}

\begin{figure*}[t]
    \centering
    \includegraphics[width=\linewidth]{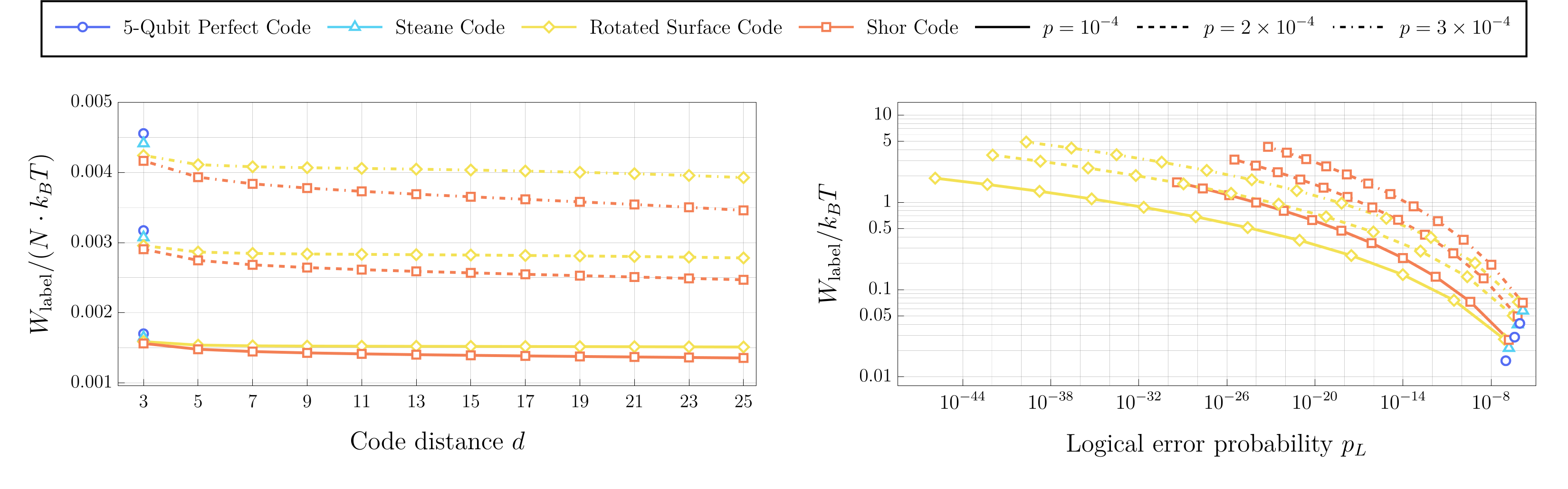}
    \caption{Erasure work associated with processed syndrome labels for selected stabiliser-code families under independent symmetric local Pauli noise with per-Pauli error probability~$p$. \textbf{Left:} Label-erasure work per physical qubit, $W_{\mathrm{label}}/(N k_{\mathrm B}T)$, versus code distance~$d$. Degeneracy reduces this quantity by merging physically distinct errors into common recovery labels. Among the families considered, this reduction is most pronounced for the Shor-type CSS codes, which consequently exhibit the lowest label-erasure work per physical qubit. \textbf{Right:} Total label-erasure work, $W_{\mathrm{label}}/(k_{\mathrm B}T)$, versus the leading-order logical-failure proxy~$p_{\mathrm L}$ defined in Appendix~\ref{app:logical_combinatorics}, for the same code families and distance range. At fixed~$p_{\mathrm L}$, rotated surface codes generally require less total label-erasure work than the corresponding Shor-type codes. Isolated markers denote the distance-three five-qubit perfect and Steane codes; at the plotted points, they require less total label-erasure work at comparable values of~$p_{\mathrm L}$, despite their larger work per physical qubit.}
    \label{fig:label_entropy}
\end{figure*}

Comparing the energetic cost of syndrome-label erasure for the Shor, surface, Steane, and five-qubit perfect codes (Eqs.~\eqref{eq:ent_perf_stean_lab} and \eqref{eq:ent_surf_lab}), we find that it depends strongly on the degeneracy structure of the underlying code. In particular, Eq.~\eqref{eq:ent_shor_lab} shows that the reduction associated with syndrome degeneracy can persist even in the large-lattice limit for Shor codes. More generally, codes whose measured stabilizer generators contain no weight-two Pauli operators exhibit no leading-order reduction in energetic cost. The Steane and five-qubit perfect codes provide examples of this behaviour, as both are defined exclusively by Pauli strings of weight four or higher [Eq.~\eqref{eq:ent_perf_stean_lab}]. 

Figure~\ref{fig:label_entropy} compares all four code families by plotting the erasure cost per physical qubit as a function of code distance and the total erasure cost as a function of logical error probability. Details of the logical error probability estimates for Shor-type and rotated surface codes are provided in Appendix~\ref{app:logical_combinatorics}. Among the codes considered, Shor codes exhibit the largest reduction in erasure cost due to syndrome degeneracy, leading to the lowest cost per physical qubit. However, this local advantage does not necessarily translate into the lowest total cost at a fixed logical error probability. In particular, the rotated surface code achieves a lower overall cost in that regime. This distinction highlights that while degeneracy in low-weight error sectors reduces the energetic cost of syndrome processing, the cost required to achieve a target logical reliability is ultimately determined by the combinatorics of higher-weight error sectors.

\section{Bit-level entropy and syndrome hierarchy}
\label{sec:bit_level_cost}

The discussion in Sec.~\ref{sec:stab_cost} treated the output of a correction cycle primarily in terms of syndrome labels, or equivalently in terms of equivalence classes of physical errors. This is the natural level of description from the perspective of the KL conditions, but it is not yet the physically primitive object generated in a measurement-based realisation of a stabiliser code. In such a realisation, the syndrome record consists of a collection of binary parity-measurement outcomes obtained from ancillary qubits coupled to the individual stabiliser generators. Any syndrome label or correction class is inferred only afterwards by classical postprocessing of this raw measurement record. This postprocessing may require a computationally intensive decoder with its own energetic cost, which we do not include. Consequently, the relevant energetic burden depends on how the syndrome information is stored and erased. In architectures that store parity outcomes locally and reset them without reversible compression, this burden is determined not only by the entropy of the final decoder output, but also by the entropy retained in the individual parity-measurement bits.

Let
$
    \mathcal{G}=\qty{ g_1,\dots,g_{r}},
$
be a choice of independent stabiliser generators to be measured, such that \mbox{$\mathcal{S} = \langle g_1,\dots,g_r\rangle$}. One round of syndrome extraction produces a binary vector
$
    \mathbf{M}=(M_1,\dots,M_r)\in\{0,1\}^r,
$
where $M_j=0$ and $M_j=1$ correspond to the measured eigenvalues $+1$ and $-1$ of the generator $g_j$, respectively. Any decoder output or compressed syndrome label is then obtained by a classical map
$
    L=f(\mathbf{M}).
$
In particular, the label-level entropy discussed in Sec.~\ref{sec:stab_cost} can only underestimate the information generated and stored in the parity-measurement. Indeed, if the decoder label is obtained by deterministic classical postprocessing, then postprocessing cannot increase Shannon entropy. Together with the subadditivity of entropy for the components of the raw record, this yields the hierarchy 
\begin{equation}\label{eq:syndrome_hierarchy} 
    H(L)\leq  H(\mathbf{M})\leq \sum_{j=1}^{r} H(M_j). 
\end{equation}
The first inequality is saturated if and only if the decoded label $L$ is uniquely determined by the raw record $\mathbf{M}$ on its support, so that no information is discarded during postprocessing. In the present error-correction setting, this occurs when all syndrome records arising with nonzero probability are assigned distinct labels. The second inequality is saturated if and only if the parity outcomes $M_1,\dots,M_r$ are statistically independent. 

Eq.~\eqref{eq:syndrome_hierarchy} defines an energetic hierarchy
which refines the earlier label-level discussion of Sec.~\ref{sec:stab_cost}. 
Following the previous convention, we introduce bit-level inefficiency as
\begin{equation} \label{eq:bit_ineff}
    \kappa_{\mathrm{bit}} \equiv \frac{\sum_{j=1}^{r} H(M_j)}{H(\mathbf{M})}
\end{equation}
which captures additional energy expenditure due to erasure of uncompressed records.
\medskip

\noindent \textbf{\textit{Single-stabiliser entropy in the ideal measurement limit}}
--
To expose the structure of the syndrome entropy at the most elementary level, consider a single stabiliser generator $g_j$, and denote by $s_j\in\{0,1\}$ the \emph{ideal} syndrome bit associated with it. If the physical noise is described by a Pauli error distribution $\{p(E)\}$, then probability $\pi_j$ of obtaining nontrivial syndrome readout corresponding to $g_j$ is given by
\begin{equation}
    P_1^{(j)}:=\Pr(s_j=1)
    =
    \sum_{E:\,\{E,g_j\}=0} p(E).
\end{equation}
In the ideal measurement limit, the ancilla readout coincides with the ideal syndrome bit, $M_j=s_j$, and the entropy generated by this single parity measurement is
\begin{equation}
        H(M_j)=H(s_j) =  h_2\qty(P_1^{(j)})
\end{equation}
where for later convenience we define binary entropy as $h_2(p) = -(1-p)\log(1-p) - p\log p$.
Thus, the single-check entropy is determined entirely by the probability that the physical error anticommutes with the measured generator.

For independent local noise models, probability $P_1^{(j)}$ can be evaluated explicitly. Suppose that $g_j$ has weight $w(g_j)$, and that each site in its support independently produces an anticommuting local event with probability~$q$. The parity measurement of $g_j$ is nontrivial iff an odd number of such local anticommuting events occur. Hence
\begin{equation}
    \begin{aligned}
        P_1^{(j)}
        & =
        \sum_{\ell\ \mathrm{odd}}
        \binom{w(g_j)}{\ell}q^\ell(1-q)^{w(g_j)-\ell} \\
        & =
        \frac{1-(1-2q)^{w(g_j)}}{2}.
    \end{aligned}
\end{equation}
For example, if $g_j$ is of $Z$-type and the local single-qubit noise has probabilities $(p_x,p_y,p_z)$, then the local anticommuting probability is $q=p_x+p_y$. For the symmetric single-qubit Pauli channel of Eq.~\eqref{eq:n_qub_noise_uniform}, this becomes $q=2p$, so that
\begin{equation}
    P_1^{(j)}
    =
    \frac{1-(1-4p)^{w(g_j)}}{2}
    =
    2p\,w(g_j)+\mathcal{O}(p^2).
\end{equation}
Therefore
\begin{equation}
    \begin{aligned}
        H(M_j)
        =
        h_2\qty(\frac{1-(1-4p)^{w(g_j)}}{2})
        =
        h_2\qty(2p\,w(g_j)+\mathcal{O}(p^2))
    \end{aligned}
\end{equation} 
Using standard expansion we find
\begin{equation}
    H(M_j)= 2p\,w(g_j)(1-\log[2p\,w(g_j)])+\mathcal{O}(p^2\log p).
\end{equation}
This shows, already at the level of a single ideal parity measurement, that high-weight checks are entropically more expensive than low-weight ones. Summing over all measured generators yields the entropy of the \emph{uncompressed} ancilla-readout stream,
\begin{equation}\label{eq:S_bit}
    H_{\mathrm{bit}}
    :=
    \sum_{j=1}^{r} h_2\!\qty(P_1^{(j)}),
\end{equation}
which is naturally interpreted as the entropy cost of an architecture in which measured parity bits are stored and reset locally, without exploiting inter-bit correlations or reversible compression. This expression can be further nuanced by introducing measurement-level noise, which we consider in Appendix~\ref{app:faulty_measurements}.

In the following three examples we examine the entropic cost of syndrome erasure bit by bit.

\begin{figure*}[t]
    \centering
   \includegraphics[width = \linewidth]{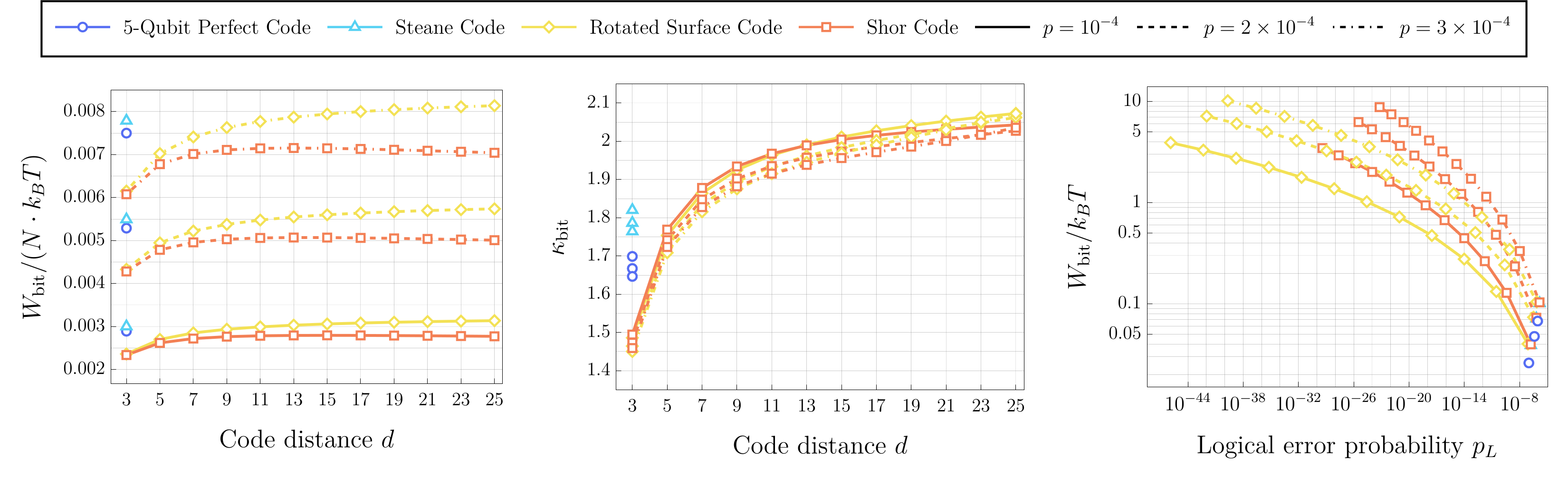}
    \caption{Erasure work associated with separately stored parity outcomes for the code families of Fig.~\ref{fig:label_entropy} under independent symmetric local Pauli noise with per-Pauli error probability~$p$. \textbf{Left:} Bit-level erasure work per physical qubit, $W_{\mathrm{bit}}/(N k_{\mathrm B}T)$, versus code distance~$d$. The surface-code value approaches its bulk-check limit, whereas the Shor-family curve reflects competition between its many weight-two checks and its $n-1$ checks of weight~$2n$, producing a finite-size maximum over the plotted range. \textbf{Centre:} Bit-level inefficiency~$\kappa_{\mathrm{bit}}$, defined in Eq.~\eqref{eq:bit_ineff}, versus code distance~$d$ for the indicated per-Pauli error probabilities. The inefficiency generally increases with distance for both scalable code families, while at small distances it decreases as~$p$ increases. \textbf{Right:} Total bit-level erasure work, $W_{\mathrm{bit}}/(k_{\mathrm B}T)$, versus the leading-order logical-failure proxy~$p_{\mathrm L}$ defined in Appendix~\ref{app:logical_combinatorics}, for the same code families and distance range. As at the label level, rotated surface codes generally require less total work than the corresponding Shor-type codes at fixed~$p_{\mathrm L}$. Isolated markers denote the distance-three five-qubit perfect and Steane codes.}
    \label{fig:bit_entropy}
\end{figure*} 

\begin{examplebox}[boxsep=0pt,left=1em,right=1em] 
\begin{center} 
	{\fontsize{10pt}{12pt}\selectfont \textbf{\hyperref[ex1]{Example 1}: Five-qubit perfect and Steane codes}} 
\end{center} 

The five-qubit perfect code and the seven-qubit Steane code contain, respectively, four and six independent stabiliser generators, all of Pauli weight~$4$. In the single-qubit error regime, their bitwise syndrome entropies are therefore 

\begin{align*} 
	H_{\text{bit},\,5\text{-q}}(p) &= 4h_2(8p)+\mathcal{O}(p^2\log p), \\ 
	H_{\text{bit},\,\mathrm{Steane}}(p) &= 6h_2(8p)+\mathcal{O}(p^2\log p). 
\end{align*} 
\end{examplebox} 

\begin{examplebox}[boxsep=0pt,left=1em,right=1em] 
\begin{center} 
	{\fontsize{10pt}{12pt}\selectfont \textbf{\hyperref[ex2]{Example 2}: $n^2$-qubit Shor codes}} 
\end{center}
 The Shor-type family contains $n(n-1)$ weight-$2$ stabilisers arising from the first repetition layer and $n-1$ stabilisers whose weight scales as $2n$. Consequently, 
\begin{equation*} 
\begin{aligned}
	H_{\text{bit},\,\mathrm{Shor}}(p) = & n(n-1)h_2(4p) \\
    &+ (n-1)h_2(4np)+\mathcal{O}(p^2\log p). 
\end{aligned}
\end{equation*} 

A meaningful large-lattice comparison requires the scalable-check probability to remain in the low-noise regime, $4np\ll 1$. Setting $p=\epsilon/n$ gives 
\begin{equation*} 
	\frac{ H_{\text{bit},\,\mathrm{Shor}}(\epsilon/n)} {n-1} = n h_2\left(\frac{4\epsilon}{n}\right) + h_2(4\epsilon). 
\end{equation*} 

Using the small-argument expansion of the binary entropy, one obtains 
\begin{equation*} 
	\frac{ H_{\text{bit},\,\mathrm{Shor}}(\epsilon/n)} {n-1} \underset{n\to\infty}{\simeq} h_2(4\epsilon) -4\epsilon \bigl[ \log(4\epsilon)-\log n-1 \bigr]. 
\end{equation*} 
Although the entropy of each individual weight-$2$ check vanishes as $n\to\infty$, their extensive multiplicity leaves a collective marginal contribution proportional to $\epsilon\log n$. 
\end{examplebox}

\begin{examplebox}[boxsep=0pt,left=1em,right=1em] 
\begin{center} 
	{\fontsize{10pt}{12pt}\selectfont \textbf{\hyperref[ex3]{Example 3}: Rotated surface codes}} 
\end{center} 

A rotated surface code of linear dimension $n$ contains $(n-1)^2$ bulk stabilisers of weight~$4$ and $2(n-1)$ edge stabilisers of weight~$2$. Its bitwise syndrome entropy is thus 
\begin{equation*} 
\begin{aligned}
	H_{\text{bit},\,\mathrm{surf}}(p) = & (n-1)^2 h_2(8p) + 2(n-1)h_2(4p) \\
    & +\mathcal{O}(p^2\log p)
\end{aligned}
\end{equation*} 

Normalising by the number $n^2-1$ of independent stabiliser measurements gives 
\begin{equation*} 
	\lim_{n\to\infty} \frac{H_{\text{bit},\,\mathrm{surf}}(p)} {n^2-1} = h_2(8p). 
\end{equation*} 

Hence, the contribution of the weight-$2$ boundary checks becomes negligible in the large-lattice limit, and the entropy per stabiliser measurement is governed by the weight-$4$ bulk. 
\end{examplebox}

We consider the same three code families as in Sec.~\ref{sec:label_track} (see boxes) and plot the resulting trends in Fig.~\ref{fig:bit_entropy}. For rotated surface codes, the growing fraction of weight-$4$ bulk checks causes the bitwise entropy per qubit to approach $h_2(8p)$ with increasing code distance, giving rise to the monotonic increase visible in the left panel. In contrast, the Shor code family combines an extensive number of low-weight checks with a smaller set of checks whose weight scales with lattice size. The interplay between these two contributions leads to a maximum at finite code size for fixed physical error probability~$p$. The central panel shows that the bitwise inefficiency $\kappa_{\mathrm{bit}}$ generally increases with code distance for both Shor and rotated surface codes at fixed~$p$. At the same time, for small code distances, $\kappa_{\mathrm{bit}}$ decreases as the physical error probability increases. The right panel compares the total bitwise entropy required to achieve a given estimated logical-error probability. As in the label-level analysis of Fig.~\ref{fig:label_entropy}, rotated surface codes generally require a lower total entropy budget than the corresponding Shor-type codes at fixed logical performance. 

From a thermodynamic perspective, these entropy trends directly determine the energetic cost of syndrome-record erasure through Landauer's principle. Consistently, all bitwise values exceed their label-level counterparts, reflecting the fact that the uncompressed syndrome record retains the entropy of individual stabilizer outcomes prior to exploiting correlations, degeneracy, or decoder-side compression. Consequently, processing syndrome data at the label level can substantially reduce the associated energetic cost, although the overall cost required to reach a target logical-error probability remains governed by the interplay between syndrome statistics and code performance.

\section{Conclusions and Outlook}
\label{sec:conclusions}

Active quantum error correction may be viewed as an entropy pump: noise-induced disorder affecting protected quantum degrees of freedom is redirected into classical syndrome information. Since physical syndrome memories are finite, this information cannot be accumulated indefinitely. It must eventually be overwritten, compressed, or erased, which necessitates energetic expenditure. In this work we isolate the entropy associated with syndrome tracking as a fundamental contribution to the thermodynamic cost of active QEC, distinct from implementation-dependent gate, control, cryogenic, and other hardware-related overheads.

At the abstract level, we show that Knill and Laflamme's conditions for error correction, together with the noise model, induce an effective syndrome state. A choice of syndrome measurement basis then determines how this state is converted into a classical record, with the resulting entropy lower-bounded by the von Neumann entropy of the effective syndrome state. In the conventional stabiliser-measurement implementation considered here, the available syndrome basis is fixed by the measured Pauli checks. In this setting, the syndrome-erasure burden is strongly shaped by degeneracy: physically distinct errors that belong to the same correction sector contribute less entropy than a naive error count would suggest. Thus the thermodynamic cost of syndrome tracking is not fixed by the parameters $[[n,k,d]]$ alone, but depends on the stabiliser structure, the noise model, and the degeneracy of the code.

We also show that the relevant entropy depends crucially on how syndrome information is represented. The effective syndrome state $\boldsymbol{\gamma}$, a processed decoder label, a raw stabiliser-measurement record, and an uncompressed collection of individual parity bits generally all carry different entropies, leading to the hierarchy of energetic costs 
\begin{equation*}
    W_{\boldsymbol{\gamma}} \leq W_{\mathrm{label}} \leq W_{\mathrm{record}} \leq W_{\mathrm{bit}}.
\end{equation*}
At each level of this hierarchy, the energetic cost is shaped by a distinct interplay between noise and code structure, as illustrated in diagram~\eqref{eq:cost_hierarchy}. We understand each level as introducing an energetic inefficiency compared to its more abstract precursor.

Over multiple QEC rounds, the distinction between $W_{\mathrm{label}}$ and $W_{\mathrm{record}}$ must be formulated in terms of the entropy rate of the syndrome process. The syndrome history generally forms a temporally correlated information process, so its erasure cost is not captured by summing independent one-round contributions. Correlations between rounds can reduce the entropy of a reversibly compressed record. A time-extended treatment must therefore distinguish the entropy rate from finite-history contributions and other transient effects. 

The information that must be retained and eventually erased also depends on the internal state of the decoder. A decoder may discard much of the complete measurement history while preserving a smaller state sufficient for subsequent recovery decisions. The corresponding erasure cost must therefore be evaluated conditionally on the retained decoder state and any other available side information. This perspective connects repeated QEC with conditional erasure in the presence of side information~\cite{Rio2011}, the thermodynamics of stochastic-pattern manipulation~\cite{Garner2017}, predictive information processing~\cite{Still2012Prediction}, and stochastic-thermodynamic descriptions of sequential information reservoirs and extended computations~\cite{BaratoSeifert2014,Wolpert2019}. A systematic investigation of the distinction between complete syndrome histories, compressed records, and decoder-retained states is deferred to future work~\cite{WIPTimeExtended}.

A complementary extension is to treat the full information flow of fault-tolerant QEC. In subsystem codes, redundant-check architectures, and qLDPC-style constructions, gauge outcomes, inferred stabiliser information, and decoder-relevant correction data need not carry the same entropy~\cite{Kribs2005OQEC,Poulin2005OperatorStabilizer,HiggottBreuckmann2021GaugeFixing,Roffe2020,Bravyi2024HighThreshold,vasic2025quantum,Hillmann2025LocalizedStatistics}. Likewise, different fault-tolerant realisations of the same abstract recovery map may generate distinct auxiliary records through verification, error detection, adaptive control, or gauge processing~\cite{Chamberland2018,Chao2020FlagFTEC,Prabhu2023,Tansuwannont2023}. The relevant question is therefore not simply which additional records appear in a particular protocol, but which information must be retained, which can be compressed or discarded conditionally on the decoder state, and which represents an unavoidable contribution to the entropy rate of the full fault-tolerant process. Resolving these distinctions will require combining code-level analysis with realistic extraction circuits, hardware-specific noise models, and online decoder implementations~\cite{Barber2025}.

More broadly, the present results suggest that syndrome entropy provides a useful bridge between abstract fault-tolerance theory and physical resource accounting for quantum error correction. The thermodynamic burden of syndrome tracking is controlled jointly by code degeneracy, check structure, noise statistics, circuit architecture, and the level at which syndrome information is resolved before erasure. These dependencies identify properties that can make syndrome processing more or less energetically demanding, but they do not by themselves establish a complete relation between energetic cost and logical error rate. Developing such a relation is an important longer-term objective: it would allow the energetic burden of a code and its implementation to be assessed against the logical protection they provide, rather than as an isolated cost. Together with a fully fault-tolerant treatment of syndrome extraction and classical processing, this would place syndrome-information thermodynamics within a more complete physical resource theory of scalable quantum error correction.

\acknowledgments

We thank Diane Zink and Yutong Luo for valuable discussions and comments during preliminary phases of the project. We further thank Wen Han Png for useful comments.
The research conducted in this publication was funded by Taighde Éireann -- Research Ireland under grant numbers IRCLA/2022/3922 and GOIPD/2026/3505.

\appendix

\section{Derivation of the effective syndrome-state description}
\label{app:effective_syndrome_derivation}

This appendix gives a detailed derivation of the effective syndrome-state formula used in Eq.~\eqref{eq:eff_dens_mat}, and of the uncorrectable-sector expressions in Eqs.~\eqref{eq:uncorr_prob}, and~\eqref{eq:uncorr_prob_bound_hier}.

We begin with the Knill--Laflamme condition
\begin{equation}
    P E_i^\dagger E_j P
    =
    \alpha_{ij}P ,
\end{equation}
where $P$ is the projector onto the code space and $\boldsymbol{\alpha}$ is a positive semidefinite matrix.  Diagonalising $\boldsymbol{\alpha}$, discarding its kernel, and rescaling the nonzero modes gives a correctable error basis $\qty{F_i}_{i=1}^r$ satisfying
\begin{equation}\label{eq:diag_KL_app}
    P F_i^\dagger F_j P
    =
    \delta_{ij}P,
    \qquad
    i,j\leq r .
\end{equation}
Equivalently, in a degenerate code with
\begin{equation}
    r
    :=
    \operatorname{rank}\boldsymbol{\alpha}
    <
    \dim\boldsymbol{\alpha},
\end{equation}
only the $r$ nonzero Knill--Laflamme modes define distinct correctable syndrome subspaces.

For each correctable error operator we use the polar decomposition
\begin{equation}
    F_iP
    =
    U_i
    \sqrt{P F_i^\dagger F_iP}
    =
    U_iP .
\end{equation}
The syndrome measurement therefore projects onto the rotated copies of the code,
\begin{equation}
    P_i
    =
    U_iPU_i^\dagger .
\end{equation}
The corresponding recovery operation can be written as the compound operation with Kraus operators
\begin{equation}
    R_i
    =
    U_i^\dagger P_i\otimes\op{i}{0},
\end{equation}
acting on the system together with a clean memory register initially in $\ket{0}$. Note that this defines a proper channel only if we assume the initial memory space to be restricted down to $\ket{0}$

The choice of syndrome basis within the correctable subspace is not unique.  For any unitary $V$ acting on the correctable index space, define
\begin{equation}
    \widetilde{F}_j
    =
    \sum_{i=1}^r V_j^iF_i .
\end{equation}
Then
\begin{align}
    P\widetilde{F}_i^\dagger\widetilde{F}_jP
    &=
    \sum_{k,l}
    \qty(V_i^k)^*
    V_j^l
    P F_k^\dagger F_l P
    \nonumber \\
    &=
    \sum_{k,l}
    \qty(V_i^k)^*
    V_j^l
    \delta_{kl}P
    \nonumber \\
    &=
    \qty(VV^\dagger)_{ji}P
    =
    \delta_{ij}P .
\end{align}
Thus every orthonormal rotation of the correctable error basis gives an equally valid set of syndrome projectors.  In this sense, the Knill--Laflamme condition associates the code with the correctable error subspace
\begin{equation}
    \mathcal{F}_{\mathrm{corr}}
    =
    \operatorname{span}\qty{F_i}_{i=1}^r ,
\end{equation}
rather than with a unique list of syndrome labels.

We now compute the syndrome statistics for a noise channel whose Kraus operators lie entirely in the correctable span,
\begin{equation}
    K_a
    =
    \sum_{i=1}^r c_a^iF_i .
\end{equation}
In the rotated basis this decomposition becomes
\begin{equation}
    K_a
    =
    \sum_{j=1}^r
    \widetilde{c}_a^j
    \widetilde{F}_j,
    \qquad
    \widetilde{c}_a^j
    =
    \sum_{i=1}^r
    c_a^i
    \qty(V^\dagger)_i^j .
\end{equation}
The rotated recovery is implemented using
\begin{equation}
    \widetilde{R}_j
    =
    \widetilde{U}_j^\dagger
    \widetilde{P}_j
    \otimes
    \op{j}{0},
\end{equation}
where
\begin{equation}
    \widetilde{F}_jP
    =
    \widetilde{U}_jP,
    \qquad
    \widetilde{P}_j
    =
    \widetilde{U}_jP\widetilde{U}_j^\dagger .
\end{equation}
For an input code state $P\rho P=\rho$, the recovered state including the syndrome register is
\begin{widetext}
\begin{align}
    \mathcal{R}\circ\mathcal{E}
    (\rho\otimes\op{0})
    &=
    \sum_{a,j}
    \widetilde{U}_j^\dagger
    \widetilde{P}_j
    K_a\rho K_a^\dagger
    \widetilde{P}_j
    \widetilde{U}_j
    \otimes
    \op{j}
    \nonumber \\
    &=
    \sum_{a,j}
    \sum_{\ell,\ell'=1}^r
    \widetilde{c}_a^\ell
    \widetilde{c}_a^{\ell'\,*}
    P\widetilde{F}_j^\dagger
    \widetilde{F}_\ell
    P\rho P
    \widetilde{F}_{\ell'}^\dagger
    \widetilde{F}_jP
    \otimes
    \op{j}
    \nonumber \\
    &=
    \sum_{a,j}
    \abs{\widetilde{c}_a^j}^2
    \rho
    \otimes
    \op{j}.
\end{align}
\end{widetext}
Taking the partial trace over the system gives
\begin{equation}
    \Tr_{\mathrm{sys}}
    \bigl[
        \mathcal{R}\circ\mathcal{E}
        (\rho\otimes\op{0})
    \bigr]
    =
    \sum_j
    \qty(
        \sum_a
        \abs{\widetilde{c}_a^j}^2
    )
    \op{j}.
\end{equation}
Introduce the effective syndrome density matrix
\begin{equation}
    \boldsymbol{\gamma}_{i'i}
    :=
    \sum_a
    c_a^{i'\,*}c_a^i .
\end{equation}
Using the relation between $c_a$ and $\widetilde{c}_a$, the probability of syndrome $j$ is
\begin{equation}
    \sum_a
    \abs{\widetilde{c}_a^j}^2
    =
    \ev{V\boldsymbol{\gamma} V^\dagger}{j}.
\end{equation}
Therefore
\begin{equation}
    \Tr_{\mathrm{sys}}
    \bigl[
        \mathcal{R}\circ\mathcal{E}
        (\rho\otimes\op{0})
    \bigr]
    =
    \sum_{j=1}^r
    \ev{V\boldsymbol{\gamma} V^\dagger}{j}
    \op{j}
    =
    \mathcal{D}_V(\boldsymbol{\gamma}),
\end{equation}
which is Eq.~\eqref{eq:eff_dens_mat} of the main text.

The map $\mathcal{D}_V$ is dephasing in the measured syndrome basis.  Hence the Shannon entropy of the measured syndrome distribution satisfies
\begin{equation}
    H[\mathcal{D}_V(\boldsymbol{\gamma})]
    \geq
    S_{\mathrm{vN}}(\boldsymbol{\gamma}),
\end{equation}
with equality when $V\boldsymbol{\gamma} V^\dagger$ is diagonal.  This proves that, within the exactly correctable sector, the entropy-optimal syndrome basis is obtained by diagonalising the effective syndrome state $\boldsymbol{\gamma}$. Notably, the resulting minimum entropy can, in principle, be achieved without performing an explicit syndrome measurement: the recovery may instead be implemented coherently via a unitary whose action is defined by $W\ket{\psi_L}\ket{0} = \sum_j \qty(U_j^\dagger P_j\ket{\psi_L})\ket{j}$, extended to a complete unitary operation, followed by tracing out the ancillary system.

We next include a branch of the noise outside the correctable span.  Let
\begin{equation}
    K'_a
    =
    \sum_{k>m}
    c_a^{\prime\,k}F_k
\end{equation}
denote Kraus operators made only from uncorrectable errors.  Their overlaps with the correctable syndrome subspaces are encoded by the logical operators
\begin{equation}
    \Delta_{\ell k}
    :=
    P F_\ell^\dagger F_k P,
    \qquad
    \ell\leq r,
    \quad
    k>m .
\end{equation}
In the rotated basis,
\begin{equation}
    \widetilde{\Delta}_{jk}
    :=
    P\widetilde{F}_j^\dagger F_kP
    =
    \sum_{\ell=1}^r
    (V_j^\ell)^*
    \Delta_{\ell k}
    =
    \sum_{\ell=1}^r
    (V^\dagger)_j^\ell
    \Delta_{\ell k}.
\end{equation}
The uncorrectable contribution to the recovered state is then
\begin{widetext}
\begin{align}
    \mathcal{R}\circ\mathcal{E}_{\mathrm{unc}}
    (\rho\otimes\op{0})
    &=
    \sum_{a,j}
    \widetilde{U}_j^\dagger
    \widetilde{P}_j
    K'_a\rho K_a^{\prime\,\dagger}
    \widetilde{P}_j
    \widetilde{U}_j
    \otimes
    \op{j}
    \nonumber \\
    &=
    \sum_{a,j}
    \sum_{k,k'>m}
    c_a^{\prime\,k}
    c_a^{\prime\,k'\,*}
    P\widetilde{F}_j^\dagger F_kP
    \rho
    P F_{k'}^\dagger\widetilde{F}_jP
    \otimes
    \op{j}
    \nonumber \\
    &=
    \sum_{a,j}
    \sum_{k,k'>m}
    c_a^{\prime\,k}
    c_a^{\prime\,k'\,*}
    \widetilde{\Delta}_{jk}
    \rho
    \widetilde{\Delta}_{jk'}^\dagger
    \otimes
    \op{j}.
\end{align}
\end{widetext}
Equivalently, define
\begin{equation}
    A_{aj}^{(V)}
    :=
    \sum_{k>m}
    c_a^{\prime\,k}
    \widetilde{\Delta}_{jk}
    =
    \sum_{k>m}
    \sum_{\ell=1}^r
    c_a^{\prime\,k}
    (V^\dagger)_j^\ell
    \Delta_{\ell k}.
\end{equation}
Then the probability of obtaining the $j$-th syndrome from the uncorrectable branch is
\begin{equation}
    p_{\mathrm{unc}}(j)
    =
    \sum_a
    \Tr\!\qty[
        A_{aj}^{(V)}
        \rho
        A_{aj}^{(V)\dagger}
    ],
\end{equation}
which is Eq.~\eqref{eq:uncorr_prob}.

We now derive the norm hierarchy used in Eq.~\eqref{eq:uncorr_prob_bound_hier}.  Since $\rho$ is a density matrix,
\begin{equation}
    \Tr[
        A\rho A^\dagger
    ]
    =
    \norm{A\sqrt{\rho}}_2^2
    \leq
    \norm{A}_\infty^2
    \norm{\sqrt{\rho}}_2^2
    =
    \norm{A}_\infty^2 .
\end{equation}
Thus
\begin{equation}
    p_{\mathrm{unc}}(j)
    \leq
    \sum_a
    \norm{A_{aj}^{(V)}}_\infty^2 .
\end{equation}
For each $a,j$,
\begin{align}
    \norm{A_{aj}^{(V)}}_\infty
    &=
    \norm{
        \sum_{k>m}
        \sum_{\ell=1}^r
        c_a^{\prime\,k}
        (V^\dagger)_j^\ell
        \Delta_{\ell k}
    }_\infty
    \nonumber \\
    &\leq
    \sum_{\ell=1}^r
    \abs{(V^\dagger)_j^\ell}
    \qty(
        \sum_{k>m}
        \abs{c_a^{\prime\,k}}
        \norm{\Delta_{\ell k}}_\infty
    )
    \nonumber \\
    &\leq
    \qty[
        \sum_{\ell=1}^r
        \qty(
            \sum_{k>m}
            \abs{c_a^{\prime\,k}}
            \norm{\Delta_{\ell k}}_\infty
        )^2
    ]^{1/2}.
\end{align}
In the final line we used the fact that the $j$-th row of $V^\dagger$ has unit Euclidean norm.  Squaring and summing over $a$ gives
\begin{equation}
    p_{\mathrm{unc}}(j)
    \leq
    \sum_a
    \sum_{\ell=1}^r
    \qty(
        \sum_{k>m}
        \abs{c_a^{\prime\,k}}
        \norm{\Delta_{\ell k}}_\infty
    )^2 .
\end{equation}
A further Cauchy--Schwarz inequality gives
\begin{align}
    \qty(
        \sum_{k>m}
        \abs{c_a^{\prime\,k}}
        \norm{\Delta_{\ell k}}_\infty
    )^2
    &\leq
    \qty(
        \sum_{k>m}
        \abs{c_a^{\prime\,k}}^2
    )
    \qty(
        \sum_{k>m}
        \norm{\Delta_{\ell k}}_\infty^2
    ).
\end{align}
Therefore
\begin{align}
    p_{\mathrm{unc}}(j)
    &\leq
    \sum_a
    \sum_{\ell=1}^r
    \qty(
        \sum_{k>m}
        \abs{c_a^{\prime\,k}}
        \norm{\Delta_{\ell k}}_\infty
    )^2
    \nonumber \\
    &\leq
    \qty(
        \sum_a\sum_{k>m}
        \abs{c_a^{\prime\,k}}^2
    )
    \qty(
        \sum_{\ell=1}^r
        \sum_{k>m}
        \norm{\Delta_{\ell k}}_\infty^2
    ),
\end{align}
which is the bound stated in Eq.~\eqref{eq:uncorr_prob_bound_hier}.  The first bound keeps more of the structure of the uncorrectable noise coefficients, while the second gives a basis-independent estimate depending only on the total uncorrectable noise weight and the total size of the logical overlaps.

\section{General KL condition and entropy-orthogonality tradeoff} \label{app:general}

We begin with the most general form of the KL condition,
\begin{equation}
    P E_i^\dagger E_j P = \alpha_{ij} P,
\end{equation}
which we use as the starting point for constructing a syndrome-detection scheme. Since the operators $\{E_i\}_i$ do not by themselves define a quantum channel, they may be mixed arbitrarily as long as their linear span is preserved. Thus, for any invertible matrix $O$, we may define
\begin{equation}
    F_\mu = O_\mu{}^j E_j,
\end{equation}
which induces the transformed KL matrix
\begin{equation}
    P F_\mu^\dagger F_\nu P = (O^\dagger \alpha O)_{\mu\nu} \equiv \alpha'_{\mu\nu}.
\end{equation}

In contrast to the diagonal KL case, errors $F_\mu$ transformed under $O$ preserve $\operatorname{span}(\qty{E_i})$ need not define orthogonal syndrome sectors. Following the standard construction, one introduces projectors
\begin{equation}
    P_\mu = U_\mu P U_\mu^\dagger,
\end{equation}
where $U_\mu$ is obtained from the polar decomposition of the projected error,
\begin{equation}
    F_\mu P = U_\mu \sqrt{P F_\mu^\dagger F_\mu P}
    = \sqrt{\alpha'_{\mu\mu}}\, U_\mu P.
\end{equation}
The corresponding syndrome projectors are generally non-orthogonal, with normalized overlaps
\begin{equation}
    \frac{\Tr(P_\mu P_\nu)}{\Tr P}
    = \frac{|\alpha'_{\mu\nu}|^2}{\alpha'_{\mu\mu}\alpha'_{\nu\nu}}.
\end{equation}
As a result, one need not have $\sum_\mu P_\mu \le \mathbb I$, so the family $\{P_\mu\}_\mu$ does not in general define a valid POVM. This can be remedied by introducing the common rescaling factor
\begin{equation}
    \varphi = \norm{\sum_\mu P_\mu}_\infty^{-1},
\end{equation}
which yields the POVM
\begin{equation}
    \qty{\varphi P_\mu}_\mu \cup \qty{\mathbb{I} - \varphi \sum_\mu P_\mu \equiv M_\neg}.
\end{equation}
The additional outcome $M_\neg$ corresponds to failed or inconclusive syndrome detection.

Now assume that the actual noise channel is described by Kraus operators
\begin{equation}
    K_i = c_i{}^j E_j = c_i{}^j (O^{-1})_j{}^\mu F_\mu.
\end{equation}
Proceeding with the usual recovery construction, one finds
\begin{align}
    U_\mu^\dagger P_\mu K_i P \sqrt{\rho}
    &= (\alpha'_{\mu\mu})^{-1/2} c_i{}^j (O^{-1})_j{}^\nu
    P F_\mu^\dagger F_\nu P \sqrt{\rho} \\
    &= (\alpha'_{\mu\mu})^{-1/2} c_i{}^j (O^{-1})_j{}^\nu
    \alpha'_{\mu\nu} P \sqrt{\rho} \\
    &= \frac{(O^\dagger)_\mu{}^k \alpha_{kj} c_i{}^j}
    {\sqrt{(O^\dagger \alpha O)_{\mu\mu}}}
    P \sqrt{\rho}.
\end{align}
It follows that the probability of obtaining syndrome label $\mu$ is
\begin{equation}
    p(\mu)
    =
    \varphi\,
    \frac{\ev{O^\dagger \alpha \boldsymbol{\gamma} \alpha O}{\mu}}
         {\ev{O^\dagger \alpha O}{\mu}},
    \qquad
    p(\neg)=1-\sum_\mu p(\mu),
\end{equation}
where
\begin{equation}
    \boldsymbol{\gamma}_{jj'} := \sum_i c_i{}^j {c_i{}^{j'}}^*
\end{equation}
is the effective syndrome density matrix. Treating $\neg$ as an additional syndrome label, the total entropy is
\begin{equation}
    S[p(\mu),p(\neg)]
    =
    -\sum_\mu p(\mu)\log p(\mu)
    -p(\neg)\log p(\neg).
\end{equation}

To expose the geometry of this optimization more clearly, it is useful to restrict to $\operatorname{supp}\alpha$ and write
\begin{equation}
    O = \alpha^{-1/2}T,
\end{equation}
with $T\in GL(r,\mathbb C)$ and $r=\rank \alpha$. Denoting the columns of $T$ by $t_\mu$, define normalized vectors
\begin{equation}
    \ket{u_\mu} := \frac{\ket{t_\mu}}{\norm{t_\mu}}.
\end{equation}
Then the syndrome probabilities take the form
\begin{equation}
    p(\mu) = \varphi\, \bra{u_\mu}\alpha^{1/2}\boldsymbol{\gamma}\alpha^{1/2}\ket{u_\mu}.
\end{equation}
Introducing
\begin{equation}
    M := \alpha^{1/2}\boldsymbol{\gamma}\alpha^{1/2},
\end{equation}
as well as the frame operator
\begin{equation}\label{eq:fram_op}
    G := \sum_\mu \ket{u_\mu}\!\bra{u_\mu},
\end{equation}
one has
\begin{equation}
    \varphi = \norm{G}_\infty^{-1},
\end{equation}
and therefore
\begin{equation}\label{eq:suc_prob}
    p(\mu)=\frac{\bra{u_\mu}M\ket{u_\mu}}{\norm{G}_\infty},
    \qquad
    p(\neg)=1-\sum_\mu p(\mu).
\end{equation}

At this point it becomes clear that the total entropy combines two distinct features of the syndrome-extraction procedure: the overall probability of obtaining a decisive outcome, and the uncertainty among the decisive syndrome labels themselves. Defining
\begin{equation}
    \eta := \sum_\mu p(\mu),
\end{equation}
together with the conditional distribution
\begin{equation}
    \widetilde p_\mu := \frac{p(\mu)}{\eta},
\end{equation}
the entropy decomposes as
\begin{equation}
    H[p(\mu),p(\neg)] = h_2(\eta) + \eta H(\widetilde p).
\end{equation}
This shows that unconstrained minimization of $H[p(\mu),p(\neg)]$ is generally not a meaningful objective: the entropy can always be lowered by decreasing $\eta$, i.e. by increasing the probability of the inconclusive outcome $\neg$.

The task is therefore naturally formulated as a tradeoff problem between decisiveness and conditional certainty. Depending on which feature is operationally most relevant, one is led to several distinct variational principles.

\subsection{Maximizing decisive syndrome detection}

A first natural objective is to maximize the total probability $\eta$ of obtaining a decisive syndrome outcome. Using~\eqref{eq:fram_op} and~\eqref{eq:suc_prob}, one finds
\begin{equation}
    \eta = \frac{\Tr(MG)}{\norm{G}_\infty}.
\end{equation}
Since $G \le \norm{G}_\infty \mathbb I$, it follows immediately that
\begin{equation}
    \eta \le \Tr M.
\end{equation}
Hence the maximal decisive-detection probability is
\begin{equation}
    \eta_{\max} = \Tr M = \Tr(\alpha\boldsymbol{\gamma}).
\end{equation}
This bound is saturated precisely when $G=\mathbb I$ on the syndrome space, i.e. when the vectors $\{\ket{u_\mu}\}_\mu$ form an orthonormal basis. In terms of the original mixing matrix, this corresponds to the orthogonal gauge
\begin{equation}
    O=\alpha^{-1/2}U,
\end{equation}
with $U$ unitary. Thus, maximizing the probability of decisive syndrome detection selects the orthogonal syndrome decomposition.

If desired, one may then use the residual unitary freedom $U$ to further minimize the conditional entropy $H(\widetilde p)$.

\subsection{Entropy minimization under a failure-probability constraint}

A second possibility is to permit inconclusive outcomes, but only up to a prescribed tolerance. One may impose the admissibility condition
\begin{equation}
    p(\neg)\le \varepsilon,
    \qquad\text{equivalently}\qquad
    \eta \ge 1-\varepsilon,
\end{equation}
for some fixed $\varepsilon\in[0,1]$. Since $\eta\le \Tr M$, such a constraint is feasible only if
\begin{equation}
    1-\varepsilon \le \Tr M.
\end{equation}
Within the admissible set, one may then minimize the total entropy,
\begin{equation}
    \min_{O\in GL(r,\mathbb C)}
    H[p(\mu),p(\neg)]
    \qquad\text{subject to}\qquad
    p(\neg)\le \varepsilon,
\end{equation}
which is conceptually similar to one-shot entropy optimizations with an error tolerance. Equivalently, one may minimize the conditional entropy of the decisive labels,
\begin{equation}
    \min_{O\in GL(r,\mathbb C)}
    H(\widetilde p)
    \qquad\text{subject to}\qquad
    p(\neg)\le \varepsilon,
\end{equation}
thereby isolating syndrome distinguishability while enforcing a minimum acceptable decisive-detection rate.

\subsection{Penalty-driven tradeoff optimization}

A third possibility is to combine the two competing desiderata into a single weighted objective. Rather than imposing a hard bound on $p(\neg)$, one introduces the penalty functional
\begin{equation}
    \mathcal F_{\lambda,\beta}(O)
    :=
    \lambda\, H(\widetilde p)
    +
    \beta\,\qty(1-\eta),
\end{equation}
with nonnegative weights $\lambda,\beta \ge 0$. Here $\lambda$ controls the importance of reducing uncertainty among decisive syndrome labels, while $\beta$ penalizes the probability of an inconclusive outcome.

Minimizing $\mathcal F_{\lambda,\beta}$ continuously interpolates between the previous two perspectives. For $\beta\gg\lambda$, the optimization strongly favors large decisive-detection probability and is therefore driven toward the orthogonal gauge. Conversely, for $\lambda\gg\beta$, it prioritizes sharply peaked conditional syndrome distributions, even at the expense of a larger failure probability.

This weighted formulation is analogous to portfolio optimization: varying the pair $(\lambda,\beta)$ generates different optimal compromises between decisiveness and conditional certainty, tracing out a Pareto frontier in the plane spanned by $(\eta,H(\widetilde p))$.

\section{Higher-weight syndrome-sector expansion}\label{app:higher-weight}

This appendix gives a systematic account of how higher-weight Pauli errors contribute to the syndrome entropy of a stabiliser code. The single-qubit case admits a relatively simple description in terms of pairwise degeneracies, since products of distinct single-qubit errors directly identify weight-two normaliser elements. For higher-weight errors this simplification is lost: different pairs of errors may yield the same normaliser element, overlapping supports can lead to cancellations, and, most importantly, a single syndrome sector may receive probability mass from errors of several different weights. As a consequence, a decomposition of the entropy into independent fixed-weight contributions is not generally correct for the physical syndrome distribution. The appropriate object is instead the full syndrome sector, together with the distribution of Pauli weights inside it. We develop this viewpoint below, first in terms of exact syndrome-sector weight enumerators and then through a sequence of coarse low-noise approximations which require progressively less detailed information about the stabiliser and normaliser structure.

The natural starting point is to work directly with syndrome sectors rather than with fixed-weight equivalence classes alone. Let 
\begin{equation}
    \mathfrak{s}:\mathcal P_n\to\mathcal S 
\end{equation}
denote the syndrome map, where $\abs{\mathcal S}=2^{n-k}$ for an $[[n,k]]$ stabiliser code.

\begin{equation}
    c_{s,m} := \abs{\qty{A\in\mathcal P_{n,m}:\mathfrak{s}(A)=s}}, 
\end{equation}
 and introduce the corresponding syndrome-sector weight enumerator 

\begin{equation}
    W_s(z) := \sum_{m=0}^{n} c_{s,m}z^m . 
\end{equation}
 For the symmetric Pauli channel of eq.~\eqref{eq:n_qub_noise_uniform}, the probability of observing syndrome $s$ is then exactly 

\begin{equation}
    \begin{aligned}
        P_s(p) & = \sum_{m=0}^{n} c_{s,m} (1-3p)^{n-m}p^m \\
        & = (1-3p)^n W_s\!\qty(\frac{p}{1-3p}). 
    \end{aligned}
\end{equation}
 Thus the syndrome entropy is 

\begin{equation}
 H(\vb p_{\mathrm{synd}}) = -\sum_{s\in\mathcal S}P_s(p)\log P_s(p). 
\end{equation}
 This expression is exact, but evaluating it requires knowledge of the full set of syndrome-sector weight enumerators, and hence essentially detailed information about the stabiliser code. 
 
 In the low-noise regime, however, much less information is needed to determine the leading behaviour. For each syndrome sector define its minimum, or birth, weight by 

\begin{equation}
 \mu_s := \min\qty{m:c_{s,m}>0}, 
\end{equation}
 and the corresponding birth multiplicity by 

\begin{equation}
 a_s := c_{s,\mu_s}. 
\end{equation}
 Then 

\begin{equation}
 P_s(p) = a_s(1-3p)^{n-\mu_s}p^{\mu_s} + \mathcal O(p^{\mu_s+1}), 
\end{equation}
 and therefore 

\begin{widetext}
    \begin{equation}
     -P_s(p)\log P_s(p) = - a_s(1-3p)^{n-\mu_s}p^{\mu_s} \log\!\qty[ a_s(1-3p)^{n-\mu_s}p^{\mu_s} ] + \mathcal O(p^{\mu_s+1}\log p). 
    \end{equation}
\end{widetext}
 The leading contribution of a syndrome sector is therefore controlled by the lowest-weight errors contained in that sector. Higher-weight errors that share the same syndrome do not create independent leading entropy terms; they only modify the probability of a syndrome sector that has already appeared at lower order. 
 
 This observation suggests a useful coarse-grained description. Let 

\begin{equation}
 B_m := \abs{\qty{s\in\mathcal S:\mu_s=m}} 
\end{equation}
 be the number of syndrome sectors which first appear at weight $m$, and define the average birth multiplicity 

\begin{equation}
 \bar a_m := \frac{1}{B_m} \sum_{\mu_s=m} a_s, 
\end{equation}
 whenever $B_m\neq0$. The quantity $B_m\bar a_m$ counts the number of weight-$m$ errors whose syndromes are not already generated at smaller weight. In terms of these coarse parameters, the leading low-noise contribution from syndrome sectors born at weight $m$ may be approximated by 

\begin{equation}
     H^{\mathrm{birth}}_m \simeq - B_m\bar a_m q_m \log\!\qty(\bar a_m q_m). 
\end{equation}
where $q_m:=(1-3p)^{n-m}p^m$. Consequently, for a cutoff weight $M$, one obtains the schematic approximation 

\begin{widetext}
    \begin{equation}
     H(\vb p_{\mathrm{synd}}) \simeq  -p_0\log p_0 - \sum_{m=1}^{M} B_m\bar a_m q_m \log\!\qty(\bar a_m q_m) + \sum_{m=2}^{M} \Delta H_m^{\mathrm{shadow}},
    \label{eq:syndrome_birth_shadow_approx}
    \end{equation}
\end{widetext}
 where $p_0=(1-3p)^n$, and $\Delta H_m^{\mathrm{shadow}}$ denotes the contribution of weight-$m$ errors whose syndromes have already appeared at lower weight. 
 
 The role of the shadow term can be understood as follows. Let $P_s^{(<m)}$ denote the probability mass accumulated in syndrome sector $s$ from errors of weight strictly smaller than $m$, and let $\delta_s^{(m)}$ be the additional probability mass contributed by weight-$m$ errors to the same sector. Then, assuming $\delta_s^{(m)}\ll P_s^{(<m)}$,

\begin{widetext}
    \begin{equation}
     - \qty(P_s^{(<m)}+\delta_s^{(m)}) \log\qty(P_s^{(<m)}+\delta_s^{(m)}) = - P_s^{(<m)}\log P_s^{(<m)} - \delta_s^{(m)} \qty(\log P_s^{(<m)}+1) + \mathcal O\!\qty( \frac{(\delta_s^{(m)})^2}{P_s^{(<m)}} ). 
    \end{equation}
\end{widetext}
 Thus the entropy scale associated with such higher-weight errors is set by the lower-weight probability $P_s^{(<m)}$, not by $q_m$ alone. This is precisely why a direct sum of fixed-weight entropy contributions generally overestimates the physical syndrome entropy: it treats the error weight as if it were part of the measured syndrome. 
 
 Equivalently, if $w$ denotes the error weight and $s$ the measured syndrome, then a fixed-weight decomposition computes the entropy of the refined random variable $(w,s)$ rather than the entropy of the syndrome alone. These are related by 

\begin{equation}
 H(s) = H(w,s)-H(w|s). 
\end{equation}
 The conditional entropy $H(w|s)$ is nonzero whenever a single syndrome sector receives contributions from several different weights. This term is negligible only when the syndrome essentially determines the error weight within the regime under consideration. 
 
 In the absence of detailed information about the code, the quantities $B_m$ and $\bar a_m$ may themselves be estimated only coarsely. Since 

\begin{equation}
 N_m:=\abs{\mathcal P_{n,m}}=3^m\binom{n}{m} 
\end{equation}
 is the number of Pauli errors of weight $m$, one always has 

\begin{equation}
 B_m\leq N_m, \qquad \sum_{r=0}^{m}B_r\leq 2^{n-k}. 
\end{equation}
 A simple saturation approximation is obtained by assuming that each new error creates a new syndrome sector until the syndrome alphabet is exhausted: 

\begin{equation}
 B_m^{\mathrm{sat}} := \min\qty{ N_m,\, 2^{n-k} - \sum_{r=0}^{m-1}B_r^{\mathrm{sat}} }, \qquad B_0^{\mathrm{sat}}=1. 
\end{equation}
 Together with $\bar a_m\simeq1$, this gives the nondegenerate, alphabet-limited approximation 

\begin{equation}
 H_{\mathrm{sat}}^{(\leq M)} \simeq -p_0\log p_0 - \sum_{m=1}^{M} B_m^{\mathrm{sat}} q_m\log q_m, 
\end{equation}
 to be understood as a leading-order estimate for syndrome sectors born at each weight, not as a decomposition over all weight-$m$ errors.
 
 Alternatively, one may use a random-occupancy model as a code-agnostic baseline. Writing 

\begin{equation}
 \Sigma:=2^{n-k}, \qquad M_m:=\sum_{r=0}^{m}N_r, 
\end{equation}
 the expected number of syndrome sectors occupied by errors of weight at most $m$ is approximated by 

\begin{equation}
 O_m^{\mathrm{hash}} \simeq \Sigma \qty[ 1- \qty(1-\frac{1}{\Sigma})^{M_m} ]. 
\end{equation}
 The corresponding estimate for the number of sectors born at weight $m$ is 

\begin{equation}
 B_m^{\mathrm{hash}} := O_m^{\mathrm{hash}}-O_{m-1}^{\mathrm{hash}}. 
\end{equation}
 This model should not be interpreted as a property of any particular stabiliser code. Rather, it provides a baseline for estimating when the syndrome alphabet starts to saturate in the absence of further structural information. 
 
 Finally, it is important to distinguish entropy reduction from correctability. For weights $m\leq t$, degeneracies among guaranteed correctable errors are stabiliser-induced. Indeed, if $A$ and $B$ both have weight at most $t$ and share the same syndrome, then 

\begin{equation}
 A^\dagger B\in\mathcal N(\mathcal S), \qquad w(A^\dagger B)\leq 2t<d. 
\end{equation}
Since $d$ is the minimum weight of a nontrivial logical operator, such an element of the normaliser cannot lie in $\mathcal N(\mathcal S)\setminus\mathcal S$; it must belong to the stabiliser group $\mathcal S$. Beyond the guaranteed correctable regime, however, syndrome merging may involve nontrivial logical operators. Such merging can still reduce the syndrome entropy, but it no longer represents harmless degeneracy from the viewpoint of error correction. 

The useful lesson is therefore that higher-weight corrections should be organised by syndrome-sector birth weights rather than by fixed physical weights alone. The detailed objects are the enumerators $W_s(z)$, while the practical coarse data are the birth counts $B_m$, the average birth multiplicities $\bar a_m$, and the shadow corrections from higher-weight errors landing in already-existing syndrome sectors.

\section{Combinatorial estimates of logical error probability}
\label{app:logical_combinatorics}

Below we provide combinatorial arguments behind the estimates of logical error probability used in Figs~\ref{fig:label_entropy} and~\ref{fig:bit_entropy}.

\subsection{Rotated surface code} 
    
    In an $n\times n$ patch of the rotated surface code with parameters $[[n^2,1,n]]$, straight minimum-weight representatives of the logical operators $X_L$ and $Z_L$ consist of vertical and horizontal length-$n$ strings of Pauli operators spanning the patch, respectively. Let $r=\left\lceil\frac{n}{2}\right\rceil$.
    
    Assuming independent uniform depolarising noise, such that each qubit experiences $X$, $Y$, or $Z$ with probability $p$, the probability that an error has a specified $X$- or $Z$-component is $2p$. Consequently, the contribution from errors supported on the straight minimum logical strings is 
    \begin{equation}
    	p_L^{\mathrm{straight}} \approx 2n\binom{n}{r} \left(2p\right)^r (1-3p)^{n^2-r}.
    \end{equation}

\smallskip

\subsection{Shor-type codes}

    Shor-type codes with parameters $[[n^2,1,n]]$ are obtained by concatenating an outer length-$n$ phase-flip repetition code with length-$n$ inner bit-flip repetition codes. They exhibit two distinct lowest-weight failure modes. Let $r=\left\lceil\frac{n}{2}\right\rceil$. First, one of the $n$ inner repetition-code blocks may fail after acquiring $X$ components on $r$ of its $n$ qubits. This gives 
    $
    	n\binom{n}{r} 
    $
    distinct lowest-weight supports. Second, the outer repetition code may fail when $r$ different inner blocks each acquire odd $Z$ parity. At the lowest physical weight, this is realised by one error with a $Z$ component in each of the $r$ selected blocks. There are $\binom{n}{r}$ ways to select the blocks and $n$ possible physical locations in each selected block, giving 
    $
    	n^r\binom{n}{r} 
    $
    distinct lowest-weight supports. Assuming independent uniform depolarising noise, the total probability assigned to these lowest-weight failure modes is therefore 
    \begin{equation} 
    	p_L^{\mathrm{low}} \approx \left(n+n^r\right)\binom{n}{r} \left(2p\right)^r (1-3p)^{n^2-r}. 
    \end{equation} 

\section{Measurement-level noise and useful syndrome information}
\label{app:faulty_measurements}

Discussion presented in section~\ref{sec:bit_level_cost} assumed perfect projective ancilla readout. In practice, however, the syndrome bit inferred by the decoder is itself usually corrupted by measurement errors. A minimal and broadly applicable model is to regard the measurement of each parity bit as a binary symmetric channel. Let $s_j$ denote the ideal syndrome bit and $M_j$ the actually recorded measurement outcome. If
\begin{equation}
    \Pr(M_j\neq s_j)=\varepsilon_j,
\end{equation}
then the probability of recording a nontrivial syndrome bit becomes
\begin{equation}\label{eq:noisy_bit_probability}
    q_j:=\Pr(M_j=1)=\varepsilon_j+(1-2\varepsilon_j)\pi_j.
\end{equation}
Thus the measured bit entropy is
\begin{equation}
    H(M_j)=h_2(q_j).
\end{equation}

The measurement entropy naturally separates into useful and useless parts. Since
\begin{equation}
    H(M_j\mid s_j)=h_2(\varepsilon_j),
\end{equation}
the mutual information between the ideal syndrome bit and the measured bit is
\begin{equation}
    I(s_j:M_j)=h_2(q_j)-h_2(\varepsilon_j).
\end{equation}
Hence $H(M_j)$ quantifies the total entropy that must be erased if the bit is stored, while $I(s_j:M_j)$ measures the useful syndrome information actually acquired by the parity measurement. Measurement noise therefore increases the raw memory burden without contributing to the decoder-relevant content of the record.

\bibliography{references}

\end{document}